\documentclass[iop]{emulateapj}
\usepackage{rotating}

\shorttitle{$^{12}$CO(J = 2 -- 1) Imaging of Seyfert 1 Galaxy NGC 1097}
\shortauthors{Hsieh et al.}

\slugcomment{Accepted for publication in {\it ApJ}}

\begin{document}

\title{Physical properties of the circumnuclear starburst ring in the barred Galaxy NGC 1097}

\author{
        Pei-Ying Hsieh\altaffilmark{1,2},
        Satoki Matsushita\altaffilmark{2,7},
        Guilin Liu\altaffilmark{3},
        Paul T. P. Ho\altaffilmark{2,6},
        Nagisa Oi\altaffilmark{4}, and
        Ya-Lin Wu\altaffilmark{2,5},
\\pyhsieh@asiaa.sinica.edu.tw}

\affil{$^1$ Institute of Astrophysics, National Central University,
        No.300, Jhongda Rd., Jhongli City, Taoyuan County 32001, Taiwan, R.O.C. }
\affil{$^2$ Academia Sinica Institute of Astronomy and
        Astrophysics, P.O. Box 23-141, Taipei 10617, Taiwan, R.O.C.}
\affil{$^3$ Astronomy Department, University of Massachusetts,
        Amherst, MA 01003, USA}
\affil{$^4$ Department of Astronomy, School of Science, Graduate University for Advanced Studies,
2-21-1 Osawa, Mitaka, Tokyo 181-8588, Japan}
\affil{$^5$ Institute of Astrophysics, National Taiwan University, No. 1, Sec. 4, Roosevelt Road,
Taipei 10617, Taiwan, R.O.C.}
\affil{$^6$ Harvard-Smithsonian Center for Astrophysics, 60 Garden Street, Cambridge, MA 02138, USA}
\affil{$^7$ Joint ALMA Office, Alonso de C$\acute{\rm o}$rdova 3107, Vitacura 763 0355, Santiago, Chile}

\begin{abstract}

We report high resolution $^{12}$CO(J = 2--1), $^{13}$CO(J = 2--1),
and $^{12}$CO(J = 3--2) imaging of the Seyfert 1/starburst ring
galaxy NGC 1097 with the Submillimeter Array to study the
physical and kinematic properties of the 1-kpc circumnuclear
starburst ring. Individual star clusters as detected in the HST map of Pa$\alpha$
line emission have been used to determine the star formation rate, and are
compared with the properties of the molecular gas.
The molecular ring has been resolved into individual clumps at
GMA-scale of 200--300 pc in all three CO lines.
The intersection between the dust lanes and the starburst ring, which is
associated with orbit-crowding region, is resolved into two
physically/kinematically distinct features in the
1\farcs5$\times$1\farcs0 (105$\times$70 pc) $^{12}$CO(J = 2--1) map. The clumps associated with the
dust lanes have broader line width, higher surface gas density, and
lower star formation rate, while the narrow line clumps associated
with the starburst ring have opposite characteristics. Toomre-Q value
under unity at the radius of the ring suggests that the molecular ring
is gravitationally unstable to fragment at the scale of the GMA.
The line widths and surface density of gas mass of the clumps show
an azimuthal variation related to the large scale dynamics. The star
formation rate, on the other hand, is not significantly affected by the
dynamics, but has a correlation with the intensity ratio of $^{12}$CO
(J = 3--2) and $^{12}$CO(J = 2--1), which traces the denser gas
associated with star formation. Our resolved CO map, especially in
the orbit-crowding region, for the first time demonstrates observationally that
the physical/kinematic properties of the GMAs are affected
by the large scale bar-potential dynamics in NGC 1097.

\end{abstract}


\section{INTRODUCTION}
\label{sect-intro}

NGC 1097 [SB(s)b; \citealt{dev91}] is a nearby
(D = 14.5 Mpc; 1$\arcsec$ = 70 pc, \citealt{tully}) barred spiral
galaxy. A pair of dust lanes are located at the leading edges of
the major bar. A radio continuum image at 1.465 GHz shows faint
ridges coinciding with the dust lanes \citep{hum87}. The nucleus
is thought to be a transition object from LINER to Seyfert 1
\citep{stor03}. Detailed studies on the nucleus show
morphological and kinematic evidences of the nuclear spirals
on the order of 30 pc, and was interpreted as part of the fueling chain
to the very center \citep{fathi06,davies09,van10}.
NGC 1097 is also an IRAS bright galaxy \citep{sanders03}.
The contribution of large amount of IR flux
arise from its 1 kpc-circumnuclear starburst ring
\citep[e.g.,][]{hum87,tele81,kot00}. The starburst ring
hosts ''hot-spots`` composed with super star clusters
identified in HST images \citep{barth95}, and was suggested
to have an instantaneous burst of star formation which occurred
$\sim$ 6--7 Myr ago \citep{kot00}.

The molecular gas of NGC 1097 in the nuclear region has
been previously mapped in the dense gas tracer of HCN(J = 1--0),
low excitation lines of $^{12}$CO(J = 1--0)
and $^{12}$CO(J = 2--1) \citep[][hereafter Paper I]{koh03,hsieh08}.
These maps show a central concentration coincident with the peak of the 6-cm radio
continuum core \citep{hum87}, as well as a molecular ring
coincident with the starburst ring. A pair of molecular ridges
coincident with the dust lanes are also detected, and show
non-circular motions, possibly caused by the bar-potential
dynamics \citep[e.g.,][]{atha,athb}. The molecular ring
has a typical warm temperature ($T_{\rm K} \sim 100$ K) and
denser gas ($n_{\rm H_{2}} \sim 10^{3}$ cm$^{-3}$) consistent
with the starburst environments \citep{wild92,aalto95}.
The molecular ring exhibits a
twin-peak structure in the 4\arcsec--10\arcsec~resolution
interferometric
CO and HCN maps, where
a pair of molecular concentrations are located in the intersection
of the molecular dust lanes and
the star forming ring. Its orientation is nearly perpendicular
to the stellar bar.
The twin-peak has higher H$_{2}$ column density
than the surrounding ring, and similar features have been seen in other
barred galaxies, and can be explained by the crowding of
gas streamlines \citep[e.g.,][]{ken92, rey97,koh99}.
The gas flow gradually changes direction and migrates
toward the center of the galaxy to accumulate to form a ring
\citep{sch84,athb,piner95}.
Subsequent enhanced star formation may occur through
gravitational fragmentation stochastically \citep{elme94},
or dynamically driven by collision of molecular clouds \citep{combes85},
or alternatively originated from the shock compressed gas near the contact point
of the dust lanes and the ring \citep{boker08}. Another intriguing 
topic is thus whether the occurrence of the circumnuclear starburst ring
would prohibit or boost its nuclear
activities \citep[e.g.,][]{tele93,sco85,heckman91,ho97}.

High spatial/kinematic resolution observations of molecular lines
are essential to study the circumnuclear ring structures, since they are the
sites of star formation and respond to the large scale dynamics.
NGC 1097 is one of the best example to study the circumnuclear ring
for its typical structures of dust lanes, starburst ring, and nuclear
activities.
In order to study the physical and kinematic properties
of the starburst ring, especially in the twin-peak region,
we have now obtained higher resolution
$^{12}$CO(J = 2--1) (1\farcs5$\times$1\farcs0),
$^{13}$CO(J = 2--1) (1\farcs8$\times1\farcs4$),
and $^{12}$CO(J = 3--2) (3\farcs5$\times2\farcs1$)
maps down to 100 pc. By virtue of the high angular resolution
multi-J lines, we derive the fundamental properties of
molecular gas as well as star formation in the ring in order
to give a comprehensive view of this system.

\section{OBSERVATIONS AND DATA REDUCTION}
\label{sect-obs}
\subsection{SMA observations}
We observed NGC 1097 with the Submillimeter Array\footnote{
The SMA is a joint project between the Smithsonian
Astrophysical Observatory and the Academia Sinica Institute 
of Astronomy and Astrophysics and is funded by the
Smithsonian Institution and the Academia Sinica.} \citep{ho04}
at the summit of Mauna Kea, Hawaii. The array consists of
eight 6-m antennas. Four basic configurations of the antennas
are available. With the compact configuration, two nights of
$^{12}$CO(J = 2--1) data were obtained in 2004 (Paper I).
To achieve higher spatial resolution, we obtained two further
nights of $^{12}$CO(J = 2--1) data with the extended and the
very extended configurations in 2005. To study the excitation
of the gas, we also obtained one night of $^{12}$CO(J = 3--2)
data in 2006 with the compact configuration. All the observations
have the same phase center. We located the
phase center at the 6-cm peak of the nucleus \citep{hum87}. Detailed observational parameters,
sky conditions, system performances, and calibration sources 
are summarized in Table~\ref{t.obspar1}.

The SMA correlator processes two IF sidebands separated by
10 GHz, with $\sim$2 GHz bandwidth each. The upper/lower
sidebands are divided into slightly overlapping 24 chunks of
104 MHz width. With the advantage of this wide bandwidth of
the SMA correlator, the receivers were tuned to simultaneously
detect three CO lines in the 230 GHz band. The
$^{12}$CO(J = 2--1) line was set to be in the upper sideband, while
the $^{13}$CO(J = 2--1) and the C$^{18}$O(J = 2--1) lines
were set to be in the lower sideband. For the $^{12}$CO(J = 3--2) line,
we placed the redshifted frequency of 344.35 GHz in the upper sideband.

We calibrated the SMA data with the MIR-IDL software package.
The detailed calibration procedures of the data of compact
configuration were described in Paper I.
For the extended and very extended configurations, where Uranus
is resolved quite severely, we observed two bright quasars
(3C454.3 \& 3C111), and adopted a similar bandpass/flux
calibration method as described in Paper I. At 345 GHz, we used
both Uranus and Neptune for the flux and bandpass calibrators.

Mapping and analysis were done with the MIRIAD and NRAO AIPS packages.
The visibility data were CLEANed in AIPS by task {\tt IMAGR}.
We performed the CLEAN process to deconvolve the dirty
image to clean image. The deconvolution procedure was adopted
with the H$\ddot{\rm o}$gbom algorithm \citep{hogbom74} and the Clark algorithm \citep{clark80}.
We used a loop gain of 10$\%$ and did restrict the CLEAN area by
iterative examinings.
The CLEAN iterations were typically stopped at the 1.5$\sigma$ residual levels.
The number of CLEAN components in individual channel maps was typically about 300.

All of the 230 GHz visibility data were combined to achieve a better $uv$
coverage and sensitivity. Our 230 GHz continuum data were constructed by
averaging the line-free channels. Due to the sideband leakage at 230 GHz,
a limited bandwidth of 1.3 GHz and 0.5 GHz were obtained respectively in the
upper and lower sidebands to make a continuum image. Weak
continuum emission at 230 GHz with a peak intensity of 10
mJy beam$^{-1}$ was
detected at about 4$\sigma$ at the southern part of the ring.
The 1$\sigma$ noise level of the continuum emission at 345 GHz
averaged over the line-free bandwidth of 0.4 GHz is 6 mJy beam$^{-1}$, 
and there is an 
$\sim3\sigma$ detection in the nucleus and in the ring.
In this paper we did not subtract the continuum emission since it is too faint
as compared to the noise level in the line maps.
The spectral line data of $^{12}$CO(J = 2--1) and $^{12}$CO(J = 3--2)
were binned to 10 km s$^{-1}$ resolution. As the $^{13}$CO(J = 2--1)
emission is fainter, those data were binned to 30 km s$^{-1}$ resolution to
increase the S/N.
The C$^{18}$O(J = 2--1) line emission
was detected in the lower sideband of the 230 GHz data. However,
the leakage from the $^{12}$CO(J = 2--1) line in the upper sideband
was significant. Therefore, in this paper we will not make use of
the C$^{18}$O(J = 2--1) data for further analysis.
In this paper, we present the maps with natural
weighting. The angular resolution and rms noise level per channel
are $1\farcs5\times1\farcs0$ and 26 mJy beam$^{-1}$ (400 mK)
for the $^{12}$CO(J = 2--1), $1\farcs8\times 1\farcs4$ and 30 mJy
beam$^{-1}$ (320 mK) for the $^{13}$CO(J = 2--1) data, and
$3\farcs5\times2\farcs1$ and 35 mJy beam$^{-1}$ (47 mK)
for the $^{12}$CO(J = 3--2) data.
We used AIPS task {\tt MOMNT} to construct the integrated
intensity-weighted maps. The task would reject pixels
lower than the threshold intensity, set to be 2.5--3$\sigma$ in the
CLEANed cube after smoothing in velocity and spatial direction.
The smoothing kernels are 30 km s$^{-1}$ in the velocity
direction, and a factor of two of the synthesized beam in
the spatial direction in the maps we made.

In the following line ratio analysis, we use the maps truncated to the same
$uv$ coverage as in Sect.~\ref{sect-ratio}

\subsection{HST NICMOS Pa${\alpha}$ image}

As one of the sample galaxies in the HST NICMOS survey of nearby galaxies
(PI: Daniela Calzetti, GO: 11080), NGC 1097 was observed on November 15, 2007 by
HST equipped with NIC3 camera. NIC3 images have a
51\arcsec$\times$51\arcsec~field of view and a plate scale of
0\farcs2 with an undersampled PSF. In this survey, each observation consists
of images taken in two narrowband filters: one centered on the Pa$\alpha$ recombination line
of hydrogen (1.87~$\mu$m) (F187N), and the other on the adjacent narrow-band continuum
exposure (F190N), which provides a reliable continuum subtraction. Each set of exposures was made with a 7-position small ($<1\arcsec$ step) dithered. Exposures of 160 and 192 seconds per dither position in F187N and F190N, respectively, reach a 1$\sigma$
detection limit of 2.3$\times$10$^{-16}$~erg~s$^{-1}$~cm$^{-2}$~arcsec$^{-2}$ in the
continuum-subtracted Pa$\alpha$ image of NGC 1097.

Using the {\tt STSDAS} package of IRAF, we removed the NICMOS {\it Pedestal}
effect, masked out bad pixels and cosmic rays, and drizzled the dither
images onto a finer (0\farcs1 per pixel) grid frame. The resultant drizzled
F190N image is then scaled and subtracted from its F187N peer after
carefully aligning to the latter using foreground stars. The residual shading
effect in the pure Pa$\alpha$ image is removed by subtracting the median
of each column. This strategy works well for the NGC 1097 data which
contain relatively sparse emission features. The PSF of the final
Pa$\alpha$ image has a 0\farcs26 FWHM. Further details of the
data reduction and image processing are described in \citet{Liu10a}.

\section{RESULTS}
\label{sect-res}
\subsection{Morphologies of the molecular gas}
\label{sect-morphology}

{
The integrated intensity maps of $^{12}$CO(J = 2--1),
$^{12}$CO(J = 3--2) and $^{13}$CO(J = 2--1) lines are shown
in Figure~\ref{fig-mom0}.
The maps show a central concentration and a ring-like structure
with a radius of 700 pc ($\sim$10\arcsec).
The gas distribution of $^{12}$CO(J = 3--2) map is similar to
that of the $^{12}$CO(J = 2--1) map, where the central concentration
has a higher integrated intensity than the ring. The $^{13}$CO(J = 2--1) map,
on the other hand, shows comparable intensity between the ring and the nucleus.
Comparing with the previous observations (Paper I),
the molecular ring and the central concentration have been resolved
into individual clumps, especially for the twin-peak structure
in the molecular ring, with 
the higher resolution $^{12}$CO(J = 2--1) map.
We also show the peak brightness temperature map
of the $^{12}$CO(J = 2--1) emission in Figure~\ref{fig-mom0}
made by the AIPS task SQASH to extract the maximum
intensity along the velocity direction at each pixel.
Note that the ring and
the nucleus have comparable brightness temperatures.

The position of the AGN \citep[6-cm radio continuum core;][]{hum87},
which is assumed to be the dynamical center of the galaxy,
seems to be offset by 0\farcs7 northwest to the central peak of the
integrated CO maps.
This is also seen in our previous results
and we interpreted it as a result of the intensity weighting in
the integration (Paper I). To confirm if the dynamical center as
derived from the $^{12}$CO(J = 1--0) emission \citep{koh03} is
consistent with that of $^{12}$CO(J = 2--1), we will have
some more analysis in Sect.~\ref{sect-kinematics}.

In Figures~\ref{fig-channel1} and \ref{fig-channel2}, we show
the $^{12}$CO(J = 2--1) and $^{12}$CO(J = 3--2) channel
maps overlaid on the archival HST I-band image (F814W) to
show the gas distribution at different velocities. We show
these images at the resolution of 40 km s$^{-1}$, but we use
the resolution of 10  km s$^{-1}$ for actual analysis.
The astrometry of the HST I-band image was corrected by
the known positions of the foreground 19 stars in the catalog of
USNO-A 2.0.
After astrometry correction, we found the intensity peak of
the nucleus of HST image has a 0\farcs8 offset northeast to the position
of AGN, this may be due to extinction, or inaccurate astrometry.
We can not rule out these factors. However this offset is within
the uncertainty of the CO synthesized beam.
Both $^{12}$CO transitions show emission with
a total velocity extent of $\sim$550--600 km s$^{-1}$ at the
2$\sigma$ intensity level. The western molecular ridge/arms
(coincident with the dust lane in the optical image) joins the
southwestern part of the ring from --162 km s$^{-1}$ to 37
km s$^{-1}$, and the eastern molecular ridge joins the northeast
ring from --42 km s$^{-1}$ to 157 km s$^{-1}$. We find that the
velocity extent of these ridges is $\sim$200 km s$^{-1}$ at
the 2$\sigma$ intensity level. In between the molecular ring
and the nuclear disk, there are also extended ridge emission
connecting the nuclear disk and the ring in both lines.
This ridge emission is more significant in the $^{12}$CO(J = 2--1)
than in the $^{12}$CO(J = 3--2) map. However, this may be an
angular resolution effect.
The ridge emission looks more significant in the $^{12}$CO(J = 2--1) map,
but the ridge emission does not have enough flux density (Jy/beam)
to reach the same S/N in the high resolution image, which means it
has extended structure, i.e., the faint ridge emission in the high
resolution map is more or less resolved.

In Figure~\ref{fig-mom0-hst}, we show the $^{12}$CO(J = 2--1)
integrated intensity map overlaid on the HST I-band (F814W)
archival image to compare the optical and radio morphologies.
The Pa$\alpha$ image (near-IR) is also shown in Figure~\ref{fig-mom0-hst}.
The optical image shows a pair of spiral arms in the central
1 kpc region. The arms consist of dusty filaments and star clusters.
Two dust lanes at the leading side of the major bar connect to
the stellar arms. Dusty filaments can also be seen to be filling
the area between the spiral arms and the nucleus. The
$^{12}$CO(J = 2--1) map (1\farcs5$\times$1\farcs0)
shows a central concentration, a molecular ring and molecular
ridges with good general correspondence to the optical image.
However, there are significant differences in a detailed comparison.
Although the starburst ring looks like spiral arms in the optical image,
the molecular gas shows a more complete ring, and the molecular
ridges are seen to join the ring smoothly.
Inside of the stellar ring, the major dust lanes are offset from the
molecular ring. Assuming that the $^{12}$CO emission is faithfully
tracing the total mass, the prominent dark dust lanes are not
significant features while they join the molecular ridges as the edge
of the ring. The stellar ring or star formation activities are then
correlated with the peaks in the total mass distribution.
The central molecular concentration corresponds in general with
the stellar nucleus, but is offset to the south of the stellar light.
The nuclear molecular distribution is quite asymmetric with lots of
protrusions at the lower intensity contours, which may
correspond to gas and dust filaments which connect the
nuclear concentration to the molecular ring.

\subsection{Properties of the molecular gas}
\label{sect-clumps}

The molecular ring has been resolved into a
complex structure of compact sources immersed in a diffuse emission
with lower surface brightness as shown in Figure~\ref{fig-mom0}.
At our resolution (100 pc), the contrast between clumps
and diffuse emission is
not high.  Hence, the clumps are possibly connected to each other via
the diffuse emission.  With such a physical configuration, it is difficult
to uniquely isolate the individual clumps by some clump finding
algorithm \citep[e.g.,][]{will94}.
Moreover, the detection of clumps is dependent on the angular resolution
which is available.
In this paper, we select the peaks of the main
structures in the $^{12}$CO(J = 2--1) integrated intensity map, in order
to locate the individual clumps within the ring and the molecular ridges.
We define the clumps located outside of the 10$\arcsec$ radius as the
dust lane clumps, and those within the 10$\arcsec$ radius as the
ring clumps.

The typical size of Giant Molecular Clouds (GMCs)
is on the order of a few to few tens pc \citep{scov87}.
The size of the clumps in the ring of NGC 1097, as detected with our
synthesized beam, is at least $\sim$200 pc. We are therefore detecting
molecular clumps larger than GMCs, most likely a group of GMCs,
namely Giant Molecular Cloud Associations \citep[GMAs;][]{vog98}.
Here we still use the term ``clumps" to describe the individual peaks
at the scale of GMA. We expect that the clumps would be resolved
into individual GMCs when higher angular resolution is available.
In Table~\ref{t.clumps}, we show the quantities
measured within one synthesized beam to study the kinematic properties
with a high resolution (i.e., one synthesize beam) in the following sections.
The observed peak brightness temperatures of individual clumps are
in the range from $\sim$2 -- 8 K.

To show the general properties of the GMAs, we also measured
the physical properties integrated over their size in Table~\ref{t.vir}.
We define the area of the GMA by measuring
the number of pixels above the threshold intensity of 5$\sigma$
in the $^{12}$CO(J = 2--1) integrated intensity map.
We calculated the ``equivalent radius'' if the measured areas are modeled as 
spherical clumps. The results
are reported in Table~\ref{t.vir}
together with the resulting $M_{\rm H_{2}}$ integrated over the area.
The derived values of $M_{\rm H_2}$ are therefore larger than that
in Table~\ref{t.clumps}, which only measured the mass within one synthesized beam
at the intensity peak.
The method to derive the $M_{\rm H_{2}}$ will be described
in Sect.~\ref{sect-mass}.
Several factors are essential to consider for fair comparisons
of our GMAs with other galaxies, such as beam size, filling factor, etc.
A rough comparisons show that our GMAs in the starburst ring have
physical extent of $\sim$200--300 pc, which have similar order
with the GMAs in other galaxies \citep[e.g.][]{rand90,tosaki07,m83}.

\subsubsection{Spectra of the molecular clumps}
\label{sect-specta}

We show in Figure~\ref{fig-spectra} the spectra of the individual clumps
selected above at their peak positions (i.e., within a synthesized beam).
Most of the clumps show single gaussian profiles.
The gaussian FWHM line widths of the clumps are determined by
gaussian fitting, and are listed in Table~\ref{t.clumps}.

The line widths we observed are quadratic sums of the
intrinsic line widths and the galactic motion across
the synthesized beam, and the relation can be expressed as
\begin{equation}
\sigma_{\rm obs}^{2} = \sigma_{\rm rot}^{2} + \sigma_{\rm int}^{2},
\end{equation}
where $\sigma_{\rm obs}$ is the observed FWHM line width,
$\sigma_{\rm rot}$ is the line width due to the galactic rotation
and the radial motions within the beam, and the $\sigma_{\rm int}$ is the
intrinsic line width. The systematic galactic motion would be negligible if
the spatial resolution is small enough. We subtract $\sigma_{\rm rot}$
as derived from a circular rotation model by fitting the rotation curve
in Sect.~\ref{sect-kinematics}.
The derived intrinsic line width of the clumps are reported in Table~\ref{t.clumps}.
In general, there is $\sim$10--20\% difference between the
observed and intrinsic line widths, which suggests that
the galactic circular motions do not dominate the line broadening at this
high resolution.
However, the large ``intrinsic'' velocity dispersion thus derived by subtracting
circular motions could still be dominated by non-circular motions,
especially in the twin-peak region and in the dust lanes.
We will mention this effect in the discussion.

In Table~\ref{t.clumps}, we define the clumps based on their location
and their velocity dispersions. The clumps in the ring are further
designated by their velocity dispersion being broader or narrower
than 30 km s$^{-1}$, and named respectively as B1,..., B3, and N1,..., N11,
respectively. The clumps located at the dust lanes are named D1,..., D5.

\subsubsection{Mass of the clumps}
\label{sect-mass}

Radio interferometers have a discrete sampling of the $uv$ coverage
limited by both the shortest and longest baselines.  Here, we estimate the
effects due to the missing information.
In Figure~\ref{fig-single}, we convolved our newly combined
SMA data to match the beam size of the JCMT data (21$\arcsec$),
and overlaid the spectra to compare the flux.
The integrated $^{12}$CO(J = 2--1) fluxes of the JCMT and SMA
data are $\sim$120 K km s$^{-1}$ and $\sim$71 K km s$^{-1}$,
respectively. The integrated $^{13}$CO(J = 2--1) fluxes of the JCMT and SMA
data are $\sim$15 K km s$^{-1}$ and $\sim$7 K km s$^{-1}$
respectively. Therefore our SMA data
recover $\sim60\%$ and $\sim47\%$ of the $^{12}$CO(J = 2--1)
and $^{13}$CO(J = 2--1) fluxes measured by the JCMT, respectively.
If the missing flux
is attributed to the extended emission, then the derived fluxes for the compact
clumps will remain reliable. 
However, the spectra of the SMA data seem to have similar line profiles as that
of the JCMT. Part of the inconsistency could then possibly be due to
the uncertainty of the flux calibration, which are $15\%$ and $20\%$ for the
JCMT and SMA, respectively. 
We note therefore that the following quantities
measured from the flux will have uncertainties of at least $20\%$
from the flux calibration.

There are several ways to calculate the molecular gas mass.
The conventional  H$_{2}$/$I_{\rm CO}$ ($X_{\rm CO}$) conversion
factor is one of the methods to derive the gas mass assuming that the
molecular clouds are virialized. First, we use the $^{12}$CO emission
to calculate the molecular H$_{2}$ column density, $N_{\rm H_{2}}$, as
\begin{equation}
N_{\rm H_{2}} = X_{\rm CO} \int{{T_{\rm b}} dv} = X_{\rm CO} I_{\rm CO},
\end{equation}
where $N_{\rm H_{2}}$ for each clump is calculated by
adopting the $X_{\rm CO}$ conversion factor of 3$\times$10$^{20}$ cm$^{-2}$ (K km s$^{-1}$)$^{-1}$ \citep{sol87}, $T_{\rm b}$ is the brightness temperature, and $dv$ is the line width.
The values are
listed in Table~\ref{t.clumps}. Note that the $X_{\rm CO}$ has
wide range of 0.5--4$\times$10$^{20}$ cm$^{-2}$ (K km s$^{-1}$)$^{-1}$
\citep{young91,strong96,dame01,draine07,meier01,meier08}. We adopted
3$\times$10$^{20}$ cm$^{-2}$ (K km s$^{-1}$)$^{-1}$ to be consistent
with Paper I.
The surface molecular gas mass density
$\Sigma_{\rm H_2}$ and the molecular gas mass $M_{\rm H_2}$
are thus calculated as,

\begin{equation}
\Sigma_{\rm H_{2}} = N_{\rm H_{2}} m_{\rm H_{2}},
\end{equation}

\begin{equation}
{M}_{\rm H_{2}} = \Sigma_{\rm H_{2}} {\rm d}\Omega,
\end{equation}
where $m_{\rm H_{2}}$ and d$\Omega$ are the mass of a hydrogen molecule
and the solid angle of the integrated area, respectively.
Since this conversion factor is used for the $^{12}$CO(J = 1--0) emission,
we assume the simplest case where the ratio of
$^{12}$CO(J = 2--1)/$^{12}$CO(J = 1--0) is unity.
The H$_{2}$ mass of the clumps measured within one synthesized beam
are listed in Table~\ref{t.clumps}. We would get the
molecular gas mass ($M_{\rm gas}$) by multiplying the H$_{2}$ mass by the mean atomic weight of 1.36 of the He correction.

The $^{12}$CO(J = 2--1) is usually optically thick and only traces the
surface properties of the clouds. Optically thinner
$^{13}$CO(J = 2--1) would be a better estimator of the total column density.
We therefore calculate the average $M_{\rm H_{2}}$ in the nucleus and the ring to
compare the $M_{\rm H_{2}}$ derived from both $^{12}$CO(J = 2--1) and $^{13}$CO(J = 2--1) lines.
In the case of $^{12}$CO(J = 2--1), the average $M_{\rm H_{2}}$ of the ring is
about 1.7$\times10^{7}$ M$_{\odot}$ within one synthesized beam, and the
corresponding value for the nucleus is 3.4$\times10^{7}$ M$_{\odot}$.
However, in Paper I we derived the intensity ratio of
$^{12}$CO(J = 2--1)/$^{12}$CO(J = 1--0) is $\sim$2 for the nucleus
in the lower resolution map. Therefore the $M_{\rm H_{2}}$ of the nucleus
is possibly smaller than that derived above. 
The conventional Galactic $X_{\rm CO}$ of 3$\times10^{20}$ cm$^{-2}$
(K km s$^{-1}$)$^{-1}$ is often suggested to be overestimated in the galactic center and the
starburst environment by a factor of 2 to 5 \citep[e.g.,][]{maloney88,meier01}.
Therefore, our estimated $M_{\rm H_{2}}$ in the ring and the nucleus
might be smaller at least by a factor of 2.

Assuming the $^{13}$CO(J = 2--1) emission is optically thin, and
the $^{13}$CO/H$_{2}$ abundance of 1$\times10^{-6}$ \citep{sol79}.
We calculate the $M_{\rm H_{2}}$ with excitation temperature ($T_{\rm ex}$)
of 20 K and 50 K in LTE condition.
The $M_{\rm H_{2}}$ of the nucleus averaged over one synthesized beam
are (7.0$\pm$2.5)$\times10^{6}$ M$_{\odot}$ and (1.3$\pm$0.4)$\times10^{7}$
M$_{\odot}$ for 20 K and 50 K, respectively.
The $M_{\rm H_{2}}$ of the ring averaged over one synthesized beam
are (5.8$\pm$2.5)$\times10^{6}$ M$_{\odot}$ and (1.1$\pm$0.4)$\times10^{7}$
M$_{\odot}$ for 20 K and 50 K, respectively. With the assumption of
constant $^{13}$CO/H$_{2}$ abundance, if the $T_{\rm ex}$ is $\le$
20 K, then the conversion factor we adopted for the $^{12}$CO line
is overestimated by a factor of $\sim$3 to 5. The overestimation
is smaller (a factor of 2 to 3) if the gas is as warm as 50 K.
We measured the $^{12}$CO(J = 2--1) flux higher than 3$\sigma$
to derive the total H$_{2}$ mass of the nucleus and the ring.
The total flux of the nucleus and the ring are
490.2$\pm$59.3 Jy km s$^{-1}$ and 2902.2$\pm$476.8 Jy km s$^{-1}$,
respectively. If we adopted the intensity ratios of $^{12}$CO(J = 2--1)
and $^{12}$CO(J = 1--0) are 1.9$\pm$0.2 and 1.3$\pm$0.2
for the nucleus and the ring (Paper I), the $M_{\rm H_2}$ of the
nucleus and the ring are (1.6$\pm$0.3)$\times$10$^{8}$ M$_{\odot}$
and (1.4$\pm$0.3)$\times$10$^{9}$ M$_{\odot}$, respectively.
Thus the nucleus and the ring account for 10\% and 90\%
of the $M_{\rm H_2}$ within the 2 kpc circumnuclear region, respectively.

We calculate the virial mass of individual clumps by
\begin{equation}
M_{\rm vir} = \frac{2r\sigma_{\rm rms}^{2}}{G},
\end{equation}
where ${\it r}$ is the radius of the clump derived in
Sect.~\ref{sect-clumps}, and $\sigma_{\rm rms}$ is the
three
dimensional intrinsic root mean square velocity dispersion,
which is equal to $\sqrt{(3/8ln2)}$$\sigma_{\rm int}$.
Since we observe the $\sigma_{\rm int}$ in one dimension, we
need to multiply the observed $\sigma_{\rm int}^{2}$ by 3.
The radius is adopted with the size of the clumps assuming
that $\sigma_{\rm int}$ is isotropic.
The results are shown in
Table~\ref{t.vir}. Note that the virial mass is for GMAs not
GMCs, and we assume that the GMAs are bounded structures. The ratio of the virial mass $M_{\rm vir}$ and
$M_{\rm gas}$ are shown in Table~\ref{t.vir}.
We found that the narrow line clumps have $M_{\rm vir}$/$M_{\rm gas}$
that are more or less about unity, but are larger than unity for the broad
line and dust lane clumps.

We plot the general properties of the molecular gas mass of individual
molecular clumps in Figure~\ref{fig-clump-prop}.
In Figure~\ref{fig-clump-prop}a, the histogram
of the gas mass integrated over their size seems to be a power law with a sharp drop at the
low mass end. This is due to the sensitivity limit,
since the corresponding 3$\sigma$ mass limit is $\sim27\times10^{6}$ M$_{\odot}$ for a clump with a diameter of 3\farcs3 (the average
of the clumps in Table~\ref{t.vir}).
In Figure~\ref{fig-clump-prop}b, the FWHM intrinsic line
widths have a weak correlation with gas mass in the ring.
In Figure~\ref{fig-clump-peak}a, we show the azimuthal
variation of $\Sigma_{\rm H_{2}}$ calculated at the emission peaks of the clumps
integrated over one synthesized beam.
If we assume that NGC 1097 has a trailing spiral, then the
direction of rotation is clockwise from east (0$\degr$)
to north (90$\degr$).
The $\Sigma_{\rm H_{2}}$ in the orbit crowding region is roughly from
0\degr~to 45\degr, and from 180\degr~to 225\degr.
Note that the dust lane clumps typically have $\Sigma_{\rm H_{2}}$ similar to
the narrow line ring clumps, which are lower than the broad line
ring clumps in the orbit crowding region. The average
$\Sigma_{\rm H_{2}}$ of the narrow line ring and dust lane
clumps is $\sim1800$ M$_{\odot}$ pc$^{-2}$, and
$\sim2300$ M$_{\odot}$ pc$^{-2}$ for the broad line ring clumps.
In Figure~\ref{fig-clump-peak}b,
the velocity dispersion of clumps located at the orbit crowding
region and the dust lane are larger than that of the narrow line
ring clumps. The average velocity dispersion of the narrow line
clumps is $\sim50$ km s$^{-1}$, and $\sim90$ km s$^{-1}$
for the broad line ring/dust lane clumps.
Given the similar line brightnesses (Table~\ref{t.clumps}) of the peaks, the increased
$\Sigma_{\rm H_{2}}$ are probably the results of increased line widths as it is
indicated in equation (2). 
On the other hand, since the $N_{\rm H_2}$
is proportional to the number density of gas and line-of-sight path (i.e., optical depth),
the higher $\Sigma_{\rm H_2}$ of the broad line clumps may be due
to either a larger number density, or a larger line-of-sight path. We will discuss
these effects in the discussion.

\subsection{Young star clusters}
\label{sect-sfr}

To check how the star formation properties are associated with the molecular
clumps in the ring, we compare the 6-cm radio
continuum, V-band, and Pa${\alpha}$ images with the molecular gas images.
The 6-cm radio continuum sources are selected from
the intensity peaks \citep{hum87}
at an angular resolution of 2\farcs5. The V-band
clusters ($<$ 13 mag) are selected from
the HST F555W image \citep{barth95}.
These V-band selected clusters have typical size of 2 pc (0\farcs03),
and are suspected to be super star clusters by \citet{barth95}.
Pa$\alpha$ clusters are identified in our HST F187N image
described in the Sect.~\ref{sect-obs}.
The Pa$\alpha$ clusters are identified by SExtractor with
main parameters of detection threshold of 15$\sigma$ ({\tt DETECT\_THRESH})
and minimum number of pixels above threshold of 1 ({\tt DETECT\_MINAREA}). The
number of deblending sub-thresholds is 50 for {\tt DEBLEND\_NTHRESH}
and 0.0005 for {\tt DEBLEND\_MINCONT}. The parameters
were chosen by a wide range of verifying and visual inspection.

We used 6-cm radio continuum and V-band selected clusters for
the phenomenological comparison with the molecular clumps.
To get a high resolution and uniform sample of star formation
rate (SFR), we use the Pa$\alpha$ clusters in the following analysis.

In the $^{12}$CO(J = 2--1) integrated intensity map (Figure~\ref{fig-mom0}),
the star clusters and radio continuum sources are located
in the vicinity of molecular clumps within the synthesized beam.
The distribution of the massive star clusters is uniform in the ring instead
of showing clustering in certain regions. The star clusters
do not coincide with most of the CO peaks.
The spatial correlation seems to be better in the peak brightness
temperature map in Figure~\ref{fig-mom0}. Furthermore, there are no detected
star clusters and radio continuum sources in the dust lane
clumps, namely, clumps D1, D2, D3, and D4.

We corrected the extinction of the Pa$\alpha$ emission by the intensity ratio
of H$\alpha$ (CTIO 1.5 m archived image) and Pa$\alpha$. The PSFs of H$\alpha$ and Pa$\alpha$
are $\sim$1\farcs0 and $\sim$0\farcs3, respectively, and an additional
convolving Gaussian kernel has been applied to both H$\alpha$
and Pa$\alpha$ images to match the CO beam size.
The observed intensity ratio of H$\alpha$ and Pa$\alpha$,
(I$_{\rm H\alpha}$/I$_{\rm Pa\alpha}$)$_{\rm o}$, and
the predicted intensity ratio
(I$_{\rm H\alpha}$/I$_{\rm Pa\alpha}$)$_{\rm i}$ are
multiplied by the extinction as follows:
\begin{equation}
\left (\frac{I_{\rm H\alpha}}{I_{\rm Pa\alpha}}\right)_{\rm o} =
\left (\frac{I_{\rm H\alpha}}{I_{\rm Pa\alpha}}\right)_{\rm i} \times
10^{-0.4{E(\rm B-V)}(\kappa_{\rm H{\alpha}}-\kappa_{\rm Pa\alpha})},
\end{equation}
where $E$(B--V) is the color excess, $\kappa_{\rm H\alpha}$ and $\kappa_{\rm Pa\alpha}$
are the extinction coefficients at the wavelength of H$\alpha$
and Pa$\alpha$, respectively.

The predicted intensity ratio of 8.6 is derived in the case B
recombination with temperature
and electron density of 10000 K and 10$^{4}$ cm$^{-3}$, respectively \citep{ost89}.
The extinction coefficient of H$\alpha$ and Pa$\alpha$ are
adopted from Cardelli's extinction curve \citep{card89},
with $\kappa_{\rm H\alpha}$ = 2.535 and $\kappa_{\rm Pa\alpha}$ = 0.455, respectively.
We derive the color excess $E$(B--V) using equation (6),
and derive the extinction of
Pa$\alpha$ using the following equation,
\begin{equation}
A_{\rm \lambda} = \kappa_{\rm \lambda} E(B-V),
\end{equation}
where $A_{\rm \lambda}$ is the 
extinction at wavelength $\lambda$.
We show the derived values for $A_{\rm Pa\alpha}$ and $A_{\rm V}$
in Table~\ref{t.sfr}, where $\kappa_{\rm V}$ is 3.1 at V-band.
The average extinction of the clumps at the wavelength of 1.88 $\micron$ (redshifted
Pa$\alpha$ wavelength) is about 0.6 mag, which is quite transparent
as compared with the extinction at the H$\alpha$ line of $\sim$ 4 mag.
We calculate SFRs using the Pa${\alpha}$ luminosity based on the
equation in \citet{cal08} of
\begin{equation}
{\rm SFR}  (\rm M_{\odot} ~\rm yr^{-1}) = 4.2\times10^{-41} L_{\rm Pa\alpha} (\rm erg ~s^{-1}).
\end{equation}

The SFR surface density ($\rm\Sigma_{SFR}$) is thus
calculated within the size of the CO synthesized beam (1\farcs5$\times$1\farcs0).
Note that SFR cannot
be determined in the dust lane clumps since we do not detect significant
star clusters in both Pa$\alpha$ and H$\alpha$ images.
The low $\rm\Sigma_{SFR}$ in the broad line ring and dust lane clumps seems to be
unlikely due to the extinction, since the extinction of the broad line
ring clumps are similar to that of the narrow line ring clumps, and the
$\Sigma_{\rm H_{2}}$ of the dust lane clumps are similar to
that of the narrow line ring clumps.
To compare the star formation activities with the properties of the
molecular gas, we measured the $\rm\Sigma_{SFR}$ at the position
of each clump (Table~\ref{t.sfr}). We show the correlation
of $\rm\Sigma_{\rm SFR}$ and the $\rm\Sigma_{H_{2}}$ of the molecular clumps in
Figure~\ref{fig-clump-sfr}a. 
This plot shows very little correlation.
In Figure~\ref{fig-kslaw}, we overlay our data on the plot of $\rm\Sigma_{SFR}$
and $\rm\Sigma_{H_{2}}$ used in Kennicutt (1998) to
compare the small and the large scale star formation.
The average number follows the Kennicutt-Schmidt correlation closely.
However, we have either lower $\rm\Sigma_{\rm SFR}$ or higher $\rm\Sigma_{H_{2}}$
than the global values in \citet{kenni98}. This might be because our spatial resolution is smaller
than for their data.
Recent investigations have shown that the power scaling
relationship of the spatially-resolved Schmidt-Kennicutt law
remains valid in the sub-kpc scale \citep{bigiel08}, to $\sim$200 pc in
M51 \citep{Liu10b} and M33 \citep{Verley10,bigiel10}, but becomes
invalid at the scale of GMC/GMAs \citep{Onodera10} because the scaling is overcome
by the large scatter. The absence of a correlation in our
100 pc study is thus not surprising because even if the
Schmidt-Kennicutt law is still valid, the scatter is expected
to as large as $\sim$0.7 dex \citep{Liu10b}, larger than
the dynamical range of the gas density in Figure~\ref{fig-clump-sfr}a.
Other possibilities to explain the inconsistence are the
uncertain conversion factor mentioned in Sect.~\ref{sect-mass}.
The conversion factor is likely to be overestimated in the galactic
center and starburst region. Moreover, regarding to the
Schmidt-Kennicutt law was derived from the global galaxies
that might be dominated by disk GMCs, our nuclear ring
might not follow the same relation for its particular
physical conditions.

As for the distribution of $\rm\Sigma_{SFR}$ in the ring,
we find that $\rm\Sigma_{SFR}$ is low in the
broad line clumps in Figure~\ref{fig-clump-sfr}b, but do not have
an obvious systematic azimuthal variation as $\rm\Sigma_{H_{2}}$ or
intrinsic line width in Figure~\ref{fig-clump-peak}.
In general,
$\rm\Sigma_{SFR}$ is higher in the northern
ring than in the southern ring. This distribution, as an
average quantity, shows no strong dependence
on location within the ring.

\subsection{Physical conditions}
\label{sect-physical}

\subsubsection{Intensity ratio of multi-J CO lines}
\label{sect-ratio}

We compare the intensity ratio of the different CO lines on the same spatial
scales by restricting the data to the same $uv$-range from 7.3 k$\lambda$ to
79.6 k$\lambda$ for the $^{12}$CO(J = 2--1), $^{13}$CO(J = 2--1) and $^{12}$CO(J = 3--2)
lines. The matched beam size of all maps is 3\farcs25$\times2\farcs55$.
We corrected for the primary beam attenuation in the maps.
We measured the line intensities of individual
clumps in the $uv$-matched integrated intensity maps, and calculated the intensity
ratios in Table~\ref{t.ratio}. The $uv$-matched
low resolution maps do have some beam smearing effects on the spectra.
However, an examination of the line profiles and attempts to
correct for line smearing did not affect the derived line ratios
to within the experimental errors.

We estimate the density and temperature of the clumps
with the LVG analysis \citep{gold74} in a one-zone
model. The collision rates of CO are from \citet{flow85}
for temperatures from 10 to 250 K, and from \citet{mckee82}
for 500 to 2000 K. We assume $^{12}$CO and $^{13}$CO abundances
with respect to H$_{2}$
of 5$\times10^{-5}$ and 1$\times10^{-6}$,
with the observed velocity gradient of $\sim$ 1 km s$^{-1}$ pc$^{-1}$
of the ring.
We determined the velocity gradient in the Paper I by the PV
diagram, and it is consistent in this paper.
The average ratio of the narrow and broad line clumps are used.
Clumps N4, B1, D1, D2, D4, D5 are excluded
in the average ratio because of their large uncertainty in R$_{13}$.
Therefore the average R$_{32}$ and R$_{13}$ of the narrow line clumps
are 1.00$\pm$0.02 and 9.90$\pm$2.11, respectively.
The average R$_{32}$ and R$_{13}$ of the broad line clumps
are 0.72$\pm$0.01 and 9.55$\pm$1.56, respectively.
With the constraint of the intensity ratios within the uncertainty,
the estimated temperature and density of the narrow
line clumps are $\ge$250 K and $(4.5\pm3.5)\times10^{3}$
cm$^{-3}$. The broad line clumps have temperatures of $45\pm15$ K
and density of $(8.5\pm1.5)\times10^{2}$ cm$^{-3}$.
The predicted brightness temperature ($T_{\rm b}$) is $\sim$100 K
for the narrow line clumps and $\sim$20 K for the broad line clumps.
However, it seems to be inconsistent with the high/low $\Sigma_{\rm H_2}$
and low/high number density in the broad/narrow line clumps
if we assume a constant scale height for the clumps. The solution may be a
smaller beam filling factor for the narrow line clumps.

In Figure~\ref{fig-clump-sfr}c, the R$_{32}$ values have a positive correlation with $\rm\Sigma_{SFR}$.
In Figure~\ref{fig-clump-sfr}d, similar to $\rm\Sigma_{SFR}$,
R$_{32}$ is slightly lower in the broad line ring clumps
and does not show any systematic pattern in the azimuthal
direction.

\subsection{Kinematics}
\label{sect-kinematics}

Figure~\ref{fig-mom1}a is the intensity weighted isovelocity map of
$^{12}$CO(J = 2--1). The gas motion in the ring
appears to be dominated by circular motion, while it
shows clear non-circular motions in the $^{12}$CO(J = 1--0) map
as indicated by the S-shape nearly parallel to the
dust lanes. As we discussed in Paper I, the non-significant
non-circular motion in the $^{12}$CO(J = 2--1) maps is
perhaps because the dust lanes are not as strongly
detected in $^{12}$CO(J = 2--1) line, along with the fact
that they are closer to the edge of our primary beam, or
the non-circular motion is not prominent at the high spatial resolution.
The circumnuclear gas is in general in solid body rotation.
The velocity gradient of the blueshifted part is slightly steeper
than the redshifted part.
We also show the intensity weighted velocity dispersion map in
Figure~\ref{fig-mom1}b.
As we mentioned above, the velocity
dispersion is larger in the twin-peak region, and lower in the region
away from the twin-peak region.

The dynamical center of NGC 1097 was derived by \citet{koh03}
in their low resolution $^{12}$CO(J = 1--0) map.
With our high resolution $^{12}$CO(J = 2--1) map, we expect to
determine the dynamical center more accurately.
We use the AIPS task {\tt GAL} to determine the dynamical center. In the task {\tt GAL}, $^{12}$CO(J = 2--1) intensity-weighted velocity map (Figure~\ref{fig-mom1}) is used to fit a rotation curve. The deduced kinematic
parameters are summarized in Table 6. We use an exponential curve to fit
the area within 7$\arcsec$ in radius. The observed rotation curve and the fitted model
curve are shown in Figure~\ref{fig-gal}. From the fitted parameters, we find that
the offset ($\sim0\farcs3$) of the dynamical center with respect to the position of the AGN is still within
a fraction of the synthesized beam size. The derived $V_{\rm sys}$ has a difference of $\sim$5 km s$^{-1}$
between $^{12}$CO(J = 1--0) and $^{12}$CO(J = 2--1) data, which is less than the velocity resolution of the data. Upon examining
the channel maps of the $^{12}$CO(J = 2--1) data, we find that the peak
of the nuclear emission is
almost coincident with the position of the AGN with an offset of 0\farcs3.
We therefore conclude
that the position offset in the integrated intensity map, as mentioned
in Sect.~\ref{sect-morphology}, is due to the
asymmetric intensity distribution.

\section{DISCUSSIONS}

\subsection{Molecular ring of NGC 1097}

\subsubsection{Twin-Peak structure}

In the low resolution CO maps (Paper I, \citealt{koh03}),
NGC 1097 shows bright CO twin-peak structure arising at the intersection
of the starburst ring and the dust lanes. The $\ge$300 pc resolution
CO data show that the barred galaxies usually have a
large amount of central concentration of molecular gas \citep{sakamoto99}.
\citet{ken92} found that in several barred galaxies which host
circumnuclear rings (M101, NGC 3351, NGC 6951),
the central concentrations of molecular gas were resolved into
twin-peak structures when resolution of $\sim$200 pc is attained.
A pair of CO intensity concentrations are found in these cases,
in the circumnuclear ring, at the intersection of
the ring and the dust lane. Their orientation is
almost perpendicular to the major stellar bar.
The twin-peak structure can be
attributed to the orbit crowding of inflowing gas stream lines.
The gas flow changes from its original orbit (the so called $x_{1}$ orbit)
when it encounters the shocks, which results in a large deflection
angle and migrate to new orbit (the so called $x_{2}$ orbit). The gas then
accumulates in the family of the $x_{2}$ orbits in the
shape of a ring or nuclear spirals \citep{atha,piner95}.
Intense massive star formation would follow in the
ring/nuclear spiral once the gas becomes dense enough
to collapse \citep{elme94}.

In our 100 pc resolution CO map, the
starburst molecular ring is resolved into individual
GMAs. In the orbit crowding region, we resolve the twin-peak
into broad line clumps associated with the curved dust
lanes. The narrow line clumps are located away from the
twin-peak and are associated with star formation.
This kind of ``spectroscopic components'' were also shown
in several twin-peak galaxies at the intersection of dust lanes
and circumnuclear ring, such as NGC 1365 \citep{sakamoto07},
NGC 4151 \citep{dumas10}, NGC 6946 \citep{schinnerer07}, and
NGC 6951 \citep{koh99}. However, most of the spectra
at these intersections show blended narrow/broad line components,
which is perhaps
due to insufficient angular resolution. Our observations
for the first time spatially resolved these two components
toward the twin-peak region of NGC 1097.

It is interesting to note that the circumnuclear ring is nearly
circular at $\sim$42$\degr$ inclination, which indicates
its intrinsic elliptical shape in the galactic plane. The schematic
sketch is shown in Figure~\ref{fig-model}.
The loci of dust lanes are invoked to trace the galactic
shock wave, and their shapes are dependent
on the parameters of the barred potential. In the case of
NGC 1097, the observed dust lanes resemble the
theoretical studies \citep{athb},
with a pair of straight lanes that slightly curve inwards in the
inner ends.
These findings are consistent with the predicted
morphology from bar-driven nuclear inflow.
The physical properties of these clumps are discussed in the following subsection.

\subsubsection{The nature of the molecular clumps in the ring}
\label{sect-size}
In Figure~\ref{fig-mom0} and Table~\ref{t.clumps}, the peak brightness temperatures
of individual clumps are from 2 to 8 K. These values are lower than the
typical temperatures of molecular gas as expected in the environment of a starburst
($\sim 100$ K), and as estimated by our LVG results in Paper I and this work.
This lower brightness temperature may be due to the small beam filling factor:

\begin{equation}
f_{\rm b} = \left(\frac{\theta^{2}_{\rm s}} {\theta^{2}_{\rm s}+\theta^{2}_{\rm b}}\right) \sim  \left(\frac{\theta^{2}_{\rm s}} {\theta^{2}_{\rm b}}\right),
\end{equation}

\begin{equation}
f_{\rm b} = \left(\frac{T_{\rm b}} {T_{\rm c}}\right)
\end{equation}
where $f_{\rm b}$ is the beam filling factor, $\theta_{\rm s}$ and $\theta_{\rm b}$ are
the source size and the beam size, respectively. We assume
$\theta_{\rm s} \ll \theta_{\rm b}$. $T_{\rm b}$ is
the brightness temperature and $T_{\rm c}$ is the actual
temperature of the clouds.
We
assume here that the source size is compact and much smaller than the beam.
If we assume the LVG predicted $T_{\rm b}\sim$100 K Sect.~\ref{sect-ratio} for 
narrow line ring clumps, and an average observed $T_{\rm b}\sim$5 K
, then $f_{\rm b}\sim$ 0.05.
This suggests $\theta_{\rm s}$ is $\sim$20 pc or association of much smaller clumps
for the narrow line ring clumps.
In the case of the broad line ring clumps, with an
average observed $T_{\rm b}$ of $\sim$5 K and 
LVG predicted $T_{\rm b}$ of $\sim$20 K, then $f_{\rm b}$ is
$\sim$0.25 and $\theta_{\rm s}$ $\sim$44 pc.
The size of the broad line ring clumps
is a factor of 2 higher than the narrow line ring clumps.
Given the estimated low volume number density of the broad line clumps
in Sect.~\ref{sect-ratio}, the inconsistency between high $\Sigma_{\rm H_2}$
and low number density mentioned is possibly
due to the large line-of-sight path, or larger scale height of the broad line clumps.
If the assumption of a spherical shape of the clumps hold, the larger
size estimated above seems to be consistent with this scenario. However,
considering the higher opacity of the broad line clumps as suggested by
the high $\Sigma_{\rm H_2}$, it is likely that we are tracing the
diffuse and cold gas at the surface of GMA rather than the dense and warm gas.
On the other hand, the narrow line clumps show the opposite trends
to trace the relatively warm and dense gas.

\subsubsection{Azimuthal variation of the line widths and $\Sigma_{\rm H_{2}}$}

In Figure~\ref{fig-clump-peak}, we found the observed (and intrinsic)
line widths, and $\Sigma_{\rm H_{2}}$, show variations
along the azimuthal direction in the ring.
There are local maxima
of line widths and $\Sigma_{\rm H_{2}}$ at the position of the twin-peak at the orbit crowding regions.
Since the brightness temperature does not vary dramatically
between the emission peaks along the ring, the deduced
peaks in H$_{2}$ column density can be directly attributed to
the increased line widths.
Beam smearing is possibly important as multiple streams may
converge within a synthesized beam. However, as shown in Fig~\ref{fig-mom1}b,
enhanced line widths occur even over extended regions.
The intrinsic velocity dispersion of both the narrow and the broad line clumps cannot be thermal,
as the implied temperature would be 2000--10000 K.
In the Galactic
GMCs, non-thermal line broadening has been attributed to
turbulence from unknown mechanisms.
In the case of NGC 1097, turbulence could have been generated
by shocks of the orbit crowding
region. Shocked gas as indicated by
H$_{2}$ S(1-0) emission 
has been reported toward the twin-peak of NGC 1097
\citep{kot00}. The narrow line clumps further along
the ring may be due to subsequent dissipation of the kinetic
energy from the shock wave.
This kind of large velocity widths up to $\sim$100 km s$^{-1}$ were also observed
in NGC 6946.  With resolutions down to GMC-scale of 10 pc,
multiple components in the spectra were seen at the nuclear twin-peak by 
\citet{schinnerer07}. However, the mechanism to cause the broadened line widths
is not clear at this scale.

As noted earlier, the ``intrinsic'' line widths we derived contain not only
the random dispersion but also the velocity gradients due to non-circular motions
generated by shock fronts.  However, our observations do not have sufficient angular resolution
to resolve the locations and magnitudes of the discrete velocity jumps to be expected across the shock front \citep{draine93}.
A detailed hydrodynamical model is also needed to quantitatively predict the magnitude of the non-circular velocity gradient in the ring.

\subsection{Gravitational stabilities of the GMAs in the starburst ring}

How do the GMAs form in the starburst ring? In the ring of NGC 1097,
we consider gravitational
collapse due to Toomre instability \citep{too64}.
We estimate the Toomre
Q parameter ($\Sigma_{\rm crit}$/$\Sigma_{\rm gas}$) to see if the rotation of the ring is able
to stabilize against the fragmentation into clumps. Here the Toomre
critical density can be expressed as:

\begin{equation}
\Sigma_{\rm crit} = \alpha
\left(\frac{\kappa \sigma_{\rm int}}{3.36{\rm G}}\right),
\end{equation}

\begin{equation}
\kappa = 1.414
\left(\frac{V}{R}\right)\left(1+\frac{R}{V}\frac{dV}{dR}\right)^{0.5},
\end{equation}
where the constant $\alpha$ is unity,
G is the gravitational constant, $\kappa$ is the epicycle
frequency, $V$ is rotational velocity, and $R$ is the radius of the ring.
If  $\Sigma_{\rm H_{2}}$ exceeds $\Sigma_{\rm crit}$,
then the gas will be gravitationally unstable and collapse.
We approximate the velocity gradient d$V$/d$R$ as close to zero
in Figure~\ref{fig-gal}.
because of the flat rotation curve of the galaxy at the position of the starburst ring. Therefore $\kappa \sim$
1.414$V$/$R$. $V$ is $\sim$ 338 km s$^{-1}$ in the galactic plane
assuming an inclination of $\sim$42\degr. We find the ratio
of $\Sigma_{\rm crit}$/$\Sigma_{\rm H_{2}}$ is less than unity
in the ring ($\sim$0.6). Since $\Sigma_{\rm gas}$ consists of H$_{2}$,
HI, and metals, this ratio of 0.6 is an upper limit.
However, the HI is often absent in centers of galaxies,
and NGC 1097 also shows a HI hole in the central 1\arcmin~\citep{hi03}.
Hence the HI gas does not have an important contribution in the nuclear region.
A ratio less than unity suggests that the ring is unstable and will
fragment into clumps.

As for whether the GMA itself is gravitational bound or not, in
Sect.~\ref{sect-mass} we found that $M_{\rm vir}$/$M_{\rm gas}$
is around unity in the narrow line clumps and larger
than unity by a factor of more than 2 in the broad line clumps.
This seems to suggest that the broad line clumps
are not virialized, probably because of the larger turbulence.
However, there are several factors to reduce the ratio.
First, since we do not subtract the non-circular motion
in the broad line clumps, the observed
$M_{\rm vir}$/$M_{\rm gas}$ should be an upper limit.
We can estimate how large the non-circular motion is if we
assume the $M_{\rm vir}$/$M_{\rm gas}$ of the broad line clumps
is unity. The magnitude of the non-circular motion is
from 50 to 100 km s$^{-1}$ for the broad line clumps
under this assumption. \citet{atha} showed that
there is a correlation between the bar axial ratio
and the velocity gradient (jump) across the shock wave.
NGC 1097 has a bar axial ratio of $\sim$2.6 \citep{men07},
which indicates the maximum velocity jump is $\sim70$ km
s$^{-1}$ (Figure 12; \citealt{athb}) . However, this number is supposed
to be an upper limit since it was measured at the strongest strength
of the shock, where the strength of the shock is a
function of position relative to the nucleus. The shock
strength seems to be weaker at the intersection of the
circumnuclear ring than for the outer straight dust lanes, 
as suggested in the model.  Hence
we expect the velocity gradient caused by shock front
is smaller than $\sim70$ km s$^{-1}$, based on this correlation.
In this case, the broad line clumps still have 20\%
larger velocity dispersion than narrow line clumps.
Second, the size of the clumps
will also be a possible factor to reduce the $M_{\rm vir}$/$M_{\rm gas}$
to unity.
In Sect.~\ref{sect-size}, we estimate that the filling factor
of the broad line clumps is $\sim$0.25, and therefore the
intrinsic radius will be smaller by a factor of 0.25$^{1/2}\sim$0.5.
Third, in Sect.~\ref{sect-mass}, we point out the molecular
gas mass derived from $^{12}$CO(J = 2--1) might be
underestimated at least by a factor of $\sim$2.
These factors also can lower the $M_{\rm vir}$/$M_{\rm gas}$
to roughly unity in most of the broad line clumps and hence
the GMA could also be gravitationally bound in the broad line
clumps.

\subsection{Star formation in the ring}
\subsubsection{Extinction}

The distribution of
the massive star clusters is uniform in the ring instead
of highly clustering in certain clumps in Figure~\ref{fig-mom0}. The star clusters
do not coincide with most of the CO peaks.  This could
be due to several reasons such as extinction or the physical
nature of the star clusters. Of course, since we are comparing
the scale of star clusters (2 pc) with that of GMAs (100 pc),
we are not able to conclude the physical correlation
by their spatial distribution.
The Pa$\alpha$ is more transparent than the commonly
used H$\alpha$ through dust extinction.
However, the foreground extinction ($E$(B-V) $\sim$ 1.3) for the
Pa$\alpha$ clusters corresponds to a hydrogen column density of
6.4$\times$10$^{21}$ cm$^{-2}$ \citep{dip94}
averaged over one synthesized beam. This is
much less than the average H$_{2}$ column density for the molecular clumps
in the ring, which is $\sim8.7\times$10$^{22}$ cm$^{-2}$. Therefore,
this suggests that the detected Pa$\alpha$ clusters might be
located on the surface of the clouds instead of being embedded inside
the clumps, or else located away from the clouds. However,
it can not be ruled out that there are deeply embedded stellar
clusters in the CO peaks in this scenario.

There is also a deficiency of star clusters in the broad
line clumps associated with the dust lanes.
This again could be due to extinction although we
found it is similar in the star forming ring and the dust lanes.
However, it is interesting that there is
a spatial offset between the Spitzer 24 $\micron$ peaks and
the CO peaks, and the FIR emission in the dust lane
is intrinsically faint based on
Herschel PACS 70 and 100 $\micron$ maps \citep{sand10}.
The long wavelength IR results are less affected by
extinction, and suggest
a lack of newly formed star clusters in the dust lanes.
However, a higher resolution for the FIR observation is needed
to confirm the star formation activities in the broad line clumps.

\subsubsection{Molecular gas and star formation}

In NGC 1097, the mechanisms of the intensive star formation in the ring
are still uncertain. It could be induced by the gravitational collapse
in the ring stochastically \citep{elme94}. The
other possible mechanism is that the stars form in the downstream
of the dust lane at the conjunction of the ring, and the
star clusters continue to orbit along the ring \citep[e.g.,][]{boker08}. The major
difference is that the latter scenario will have an age gradient
for the star clusters along the ring while it is randomized in
the previous case. Several papers have discussed these mechanisms
in the galaxies that have star forming rings and there is no
clear answer so far \citep[e.g.,][]{mazz08,boker08,buta00}.
\citet{sand10} tested the above pictures by
examining if there is an azimuthal gradient of dust temperature
in the ring of NGC 1097, assuming that the younger population
of massive star clusters will heat the dust to higher temperatures
than older clusters. There seems to be no gradient, though it
is difficult to conclude since a few rounds of galactic rotation might
smooth out the age gradient.
We do not aim to solve the above question in this paper
since the most direct way is to measure the age of the star
clusters, and this needs detailed modeling. It is interesting to compare
the properties of the molecular clumps with star clusters since the
molecular clouds are the parent site of star formation.
The one relevant result here is that the line widths
are narrower further away from where the molecular
arms join the ring. This could be related to the dissipation
of turbulence which may allow cloud collapse to proceed.

In Figure~\ref{fig-clump-sfr}(b), we found that $\Sigma_{\rm SFR}$,
compared with velocity dispersion (Figure~\ref{fig-clump-peak}),
has no significant azimuthal correlation. Furthermore,
R$_{32}$  shows a similar trend to $\Sigma_{\rm SFR}$
as a function of azimuthal direction (Figure~\ref{fig-clump-sfr}(d)),
where R$_{32}$ and $\Sigma_{\rm SFR}$ has a correlation
in Figure~\ref{fig-clump-sfr}(c).
This suggests that $\Sigma_{\rm SFR}$ and R$_{32}$
are physically related, but might not be associated with the large
scale dynamics in this galaxy. Although $\Sigma_{\rm SFR}$ seems to
be suppressed in the broad line
clumps. Nevertheless, in Figure~\ref{fig-clump-sfr}(b)\&(d)
we consider that the standard deviations of the measured $\Sigma_{\rm SFR}$ do not significantly diverse among global variations. The standard deviations are 0.7 M$_{\odot}$ yr$^{-1}$ kpc$^{-2}$ and 0.4 M$_{\odot}$ yr$^{-1}$ kpc$^{-2}$ of the northeast and southwest of the ring, respectively. The mean values of the
$\Sigma_{\rm SFR}$ are 3.3 M$_{\odot}$ yr$^{-1}$ kpc$^{-2}$
and 1.4 M$_{\odot}$ yr$^{-1}$ kpc$^{-2}$ of the northeast
and southwest of the ring, respectively.
With the limited points, our results
suggest that the star formation activities are randomly generated on the local scale
instead of a systematic distribution.

In Figure~\ref{fig-lvg}, we show how the R$_{32}$ ratio varies for
different densities and kinetic temperatures.
It shows that when R$_{32}$ varies from 0.3 to 0.9, the required
number density of molecular gas changes from 10$^{2}$ cm$^{-3}$ to 5$\times10^{3}$ cm$^{-3}$.
The R$_{32}$ seems to be dependent on the density more than
the temperature when it is below unity.
Hence the variation of R$_{32}$ indicates different
density among clumps.
It is interesting to note that there is a correlation between
R$_{32}$ and SFR,
and some clumps (N1, N2, N7) which have higher SFR values
are spatially close to the HCN(J = 1--0) peaks
\citep{koh03}. Higher values of
R$_{32}$ will select denser gas associated with
higher SFR, as has been shown in the large scale observations of HCN and FIR correlation \citep{gao04}. In the smaller GMC-scale, \citet{lada92} also showed
that the efficiency of star formation is higher in the dense core rather than in the
diffuse gas.
The ratio
map of R$_{32}$ can be useful to determine
the location of star formation.

\section{SUMMARY}
\begin{enumerate} 
\item We show the multi-J CO line maps of NGC 1097
			toward the 1 kpc circumnuclear region. The molecular
			ring is resolved into individual GMAs in the
			star forming ring, the dust lanes, and at the twin-peak structures.
			For the first time the molecular concentration at the twin-peak is resolved in to
			two populations of GMAs in terms of velocity dispersion and physical conditions.
			The clumps in the starburst ring
			have narrower velocity dispersion while the line widths are
               broader in the dust lanes, and for some clumps located in the
			twin-peak. The physical and kinematic properties are
			different for these clumps.
			The narrow line clumps
			have higher temperature ($\ge$250 K) and density
			($(4.5\pm3.5)\times10^{3}$ cm$^{-3}$)
			in contrast to the broad line clumps
			(T = 45$\pm$15 K; $N_{\rm H_{2}}$ =
			$(8.5\pm1.5)\times10^{2}$ cm$^{-3}$ based on the LVG analysis.		

\item The Toomre-Q factor is smaller than unity in the molecular ring
               suggesting that the GMAs
			could form via gravitational instability in the ring,
			where the $\Sigma_{\rm H_{2}}$ of the clumps
			is large enough to overcome the critical density.
			The narrow line clumps are gravitationally
			bound as shown by the values of $M_{\rm vir}$/$M_{\rm gas}$
               which are
			nearly unity. Although $M_{\rm vir}$/$M_{\rm gas}$
			is larger than unity in the broad line clumps, by 
               accounting for non-circular motions,
			smaller intrinsic source sizes, and the underestimation of molecular
			gas mass, we can lower $M_{\rm vir}$/$M_{\rm gas}$
			to unity. Therefore both systems are likely to be
			gravitationally bound.

\item	The SFR is correlated to R$_{32}$, suggesting that
			the star formation activities and the physical conditions
			of the molecular gas are associated with each other.
			In contrast to the velocity dispersion and the
			$\Sigma_{\rm H_{2}}$, the SFR and
			R$_{32}$ are not correlated with the large scale
			dynamics. This suggests that the visible star formation
        activities remain a localized phenomenon.
			The SFR is lower in the broad line clumps
			than in the narrow line clumps, which may be intrinsicly
			suppressed in the dust lanes.
						
\end{enumerate}

\acknowledgements
We thank the SMA staff for maintaining the operation of the array.
We appreciate for referee's detail comments to improve the manuscript.
We thank G. Petitpas for providing the JCMT data.
P.-Y. Hsieh especially acknowledges the fruitful discussions with L.-H. Lin,
K. Sakamoto, N. Scoville, W. Maciejewski, and L. Ho for the manuscript.
This project is funded by NSC 97-2112-M-001-007-MY3
and NSC 97-2112-M-001-021-MY3.

\acknowledgements

{\it Facilities:} \facility{SMA}, \facility{HST (NICMOS)}

\clearpage


\begin{center}
\begin{deluxetable}{lcccc}
\tablewidth{0pt}
\tabletypesize{\scriptsize}
\setlength{\tabcolsep}{0.02in} 
\tablecaption{SMA observation parameters \label{t.obspar1}}
\tablehead{
\multicolumn{1}{c}{} & {230 GHz} &
\colhead{230 GHz} &
\colhead{230 GHz} & {345 GHz}
}

\startdata
\multicolumn{1}{c}{Parameters} & {Compact-N} &
\colhead{Extended} &
\colhead{Very extended} & {Compact}\\
\hline \\
Date & 2004-07-23, 2004-10-01 & 2005-09-25 & 2005-11-07 & 2006-09-05\\
Phase center (J2000.0):\\
\multicolumn{1}{c}{R.A.} & \multicolumn{4}{c}{$\alpha_{2000}$ = 02$^{\rm h}$46$^{\rm m}$18$\fs$96}\\
\multicolumn{1}{c}{Decl.} & \multicolumn{4}{c}{$\delta_{2000}$ = --30${\degr}$16${\arcmin}$28 $
{\farcs}$897}\\
Primary beams & \multicolumn{3}{c}{52$\arcsec$} & {36$\arcsec$}\\
No. of antennas & 8, 8 & 6 & 7 & 7 \\
Project baseline range (k$\lambda$) & 5 -- 74, 10 -- 84 & 23 -- 121 & 12 -- 390 & 7.2 -- 80\\
Bandwidth (GHz)& \multicolumn{4}{c}{1.989}\\
Spectral resolution (MHz) &0.8125, 3.25 & 0.8125 & 0.8125 & 0.8125\\
Central frequency, LSB/USB (GHz) & \multicolumn{3}{c}{219/228} &{334/344}\\
$\tau_{225}$\tablenotemark{b} & 0.15, 0.3 & 0.06 & 0.1 & 0.06\\
T$_{\rm sys,DSB}$ (K) &200, 350 & 110 & 180 & 300\\
Bandpass calibrators & Uranus, J0423 -- 013 & 3C454.3 & 3C454.3, 3C111 & Uranus, Neptune\\
Absolute flux calibrators\tablenotemark{a} & Uranus (36.8, 34.2),& Uranus (37.2), & Uranus (34.9), & Neptune (21.1)\\
& J0423 -- 013 (2.8, 2.5)& 3C454.3 (21.3) & 3C454.3 (21.3) & \\
Gain calibrators & J0132 -- 169 & J0132 -- 169 & J0132 -- 169 & J0132 -- 169, J0423 -- 013\\

\enddata

\tablenotetext{a}{
The numbers in the parenthesis are the absolute flux in Jy.}
\tablenotetext{b}{$\tau_{225}$ is the optical depth measured in 225 GHz.}

\end{deluxetable}
\end{center}
\clearpage

\begin{deluxetable}{cccccccccccccc}
\tablewidth{0pt}
\tabletypesize{\scriptsize}
\setlength{\tabcolsep}{0.02in}
\tablecaption{Physical parameters of the peaks of the molecular clouds \label{t.clumps}}
\tablehead{\colhead{ID} & \colhead{$\delta$R.A.} & \colhead{$\delta$Decl.} & \colhead{$I_{\rm CO}$} &
\colhead{$\delta$$V_{\rm obs}$} & \colhead{$\delta$V$_{\rm int}$} & \colhead{$N_{\rm H_2}$} &
\colhead{$\Sigma_{\rm H_2}$} & \colhead{$M_{\rm H_2}$} & \colhead{$T_{\rm b}$}\\
\colhead{} & \colhead{(1)} & \colhead{(2)} & \colhead{(3)} & \colhead{(4)} & \colhead{(5)} &  
\colhead{(6)} & \colhead{(7)} & \colhead{(8)} & \colhead{(9)}\\
\colhead{} & \colhead{(\arcsec)} & \colhead{(\arcsec)} & \colhead{(Jy beam$^{-1}$ km s$^{-1}$)} & \colhead{(km s$^{-1}$)} & \colhead{(km s$^{-1}$)} &  
\colhead{(10$^{22}$ cm$^{-2}$)} & \colhead{(M$_{\odot}$ pc$^{-2}$})
& \colhead{(10$^{6}$ M$_{\odot}$)}
& \colhead{(K)}
}

\startdata
N1	&	9.0	&	1.6	&	17.2	&	54$\pm$3	&	42$\pm$4	&	7.9	&	1270	&	10.7	&	4.1	\\
N2	&	8.6	&	4.8	&	27.9	&	67$\pm$3	&	61$\pm$4	&	12.8	&	2070	&	17.3	&	5.5	\\
N3	&	3.6	&	8.4	&	38.3	&	69$\pm$1	&	64$\pm$1	&	17.6	&	2840	&	23.8	&	7.5	\\
N4	&	0.6	&	8.2	&	18.5	&	62$\pm$4	&	56$\pm$4	&	8.5	&	1370	&	11.5	&	5.8	\\
N5	&	-4.8	&	7.8	&	21.3	&	38$\pm$1	&	31$\pm$1	&	9.8	&	1580	&	13.2	&	6.1	\\
N6	&	-6.4	&	5.6	&	19.6	&	41$\pm$2	&	37$\pm$2	&	9.0	&	1460	&	12.2	&	4.1	\\
N7	&	-8.6	&	-2.6	&	40.3	&	61$\pm$2	&	52$\pm$3	&	18.5	&	2990	&	25.0	&	8.0	\\
N8	&	-5.4	&	-8.4	&	28.4	&	57$\pm$2	&	51$\pm$2	&	13.1	&	2110	&	17.6	&	5.8	\\
N9	&	-2.8	&	-9.2	&	24.6	&	62$\pm$3	&	57$\pm$3	&	11.3	&	1820	&	15.2	&	4.1	\\
N10	&	4.8	&	-7.8	&	19.7	&	43$\pm$2	&	37$\pm$2	&	9.1	&	1460	&	12.2	&	6.1	\\
N11	&	7.8	&	-6.6	&	27.7	&	52$\pm$4	&	49$\pm$4	&	12.7	&	2050	&	17.2	&	5.8	\\
B1	&	10.2	&	3.6	&	27.8	&	113$\pm$10	&	109$\pm$11	&	12.7	&	2060	&	17.2	&	2.4	\\
B2	&	-7.4	&	-6.8	&	34.8	&	99$\pm$5	&	96$\pm$5	&	16.0	&	2580	&	21.6	&	4.6	\\
B3	&	-5.4	&	-6.6	&	31.3	&	94$\pm$8	&	90$\pm$8	&	14.4	&	2320	&	19.4	&	4.1	\\
D1	&	16.8	&	-1.4	&	18.9	&	100$\pm$10	&	97$\pm$10	&	8.7	&	1400	&	11.7	&	1.8	\\
D2	&	12.8	&	1.6	&	21.1	&	86$\pm$17	&	82$\pm$18	&	9.7	&	1560	&	13.1	&	2.0	\\
D3	&	-17.8	&	2.6	&	16.2	&	81$\pm$7	&	78	$\pm$	8	&	7.5	&	1200	&	10.1	&	1.8	\\
D4	&	-13.0	&	-1.0	&	24.4	&	84$\pm$5 &	80$\pm$6	&	11.2	&	1810	&	15.1	&	3.5	\\
D5	&	-10.0	&	-4.0	&	38.6	&	118$\pm$4	&	115$\pm$4	&	17.7	&	2860	&	24.0	&	4.6	\\
Nu	&	0	&	0	&	50.4	&		52$\pm$5		&	-	&	16.1	&	3740	&	23.2	&	-	\\
Nu	&	0	&	0	&	-	&		57$\pm$5		&-		&	-	&	-	&	-	&	-	\\
Nu	&	0	&	0	&	-	&		186$\pm$20		&-	  &-		&-	 &-		&	-	\\
\enddata

\tablecomments{
We define the peaks based on their location and their
velocity dispersions. The clumps in the dust lanes (molecular
spiral arms) are named D1, ..., D5. The clumps in the ring
are further designated by their velocity dispersion being
broader or narrower than 30 km s$^{-1}$, and named respectively
as B1, ..., B3, and N1, ..., N11. Nu is the ID of the nucleus.
(1) R.A. offsets from the phase center.
(2) Dec. offsets from the phase center.
(3) Integrated CO(J = 2 -- 1) intensity. The uncertainty is 2.2 Jy beam$^{-1}$ km s$^{-1}$.
(4) Fitted FWHM for the observed line width.
The nucleus has multiple-gaussians profile, and
we list the fitted line widths with 3 gaussians. Note that the $I_{\rm CO}$,
$N_{\rm H_{2}}$, and $\Sigma_{\rm H_2}$ of the nucleus
is the sum value of the three components. 
(5) FWHM line width of intrinsic velocity dispersion.
(6) H$_{2}$ column density with the uncertainty of 1.1$\times$10$^{22}$ cm$^{-2}$.
(7) $\Sigma_{\rm H_{2}}$ with the uncertainty of 170 M$_{\odot}$ pc$^{-2}$.
(8) Mass of molecular H$_{2}$ within the synthesized beam (1\farcs5$\times$1\farcs0). The uncertainty is 1.4
$\times$10$^{6}$ M$_{\odot}$.
(9) Peak brightness temperature.}

\end{deluxetable}

\clearpage

\begin{deluxetable}{ccccccc}
\tablewidth{0pt}
\tabletypesize{\scriptsize}
\tablecaption{Physical parameters of the GMAs\label{t.vir}}
\tablehead{\colhead{ID} & 
\colhead{Diameter}&
\colhead{$M_{\rm H_{2}}$} &
\colhead{$M_{\rm gas}$} &
\colhead{$M_{\rm vir}$} &
\colhead{$M_{\rm vir}$/$M_{\rm gas}$}\\
\colhead{} & \colhead{(1)} & \colhead{(2)} & \colhead{(3)} &
\colhead{(4)} & \colhead{(5)}\\
\colhead{} & \colhead{(\arcsec)} & \colhead{(10$^{6}$ M$_{\odot}$)} & \colhead{(10$^{6}$ M$_{\odot}$)} &  \colhead{(10$^{6}$ M$_{\odot}$)} & \colhead{}
}

\startdata
N1	&	2.9	&	32.4	$\pm$	5.3	&	44.1	$\pm$	7.2	&	44.9	$\pm$	10.2	&	1.0	$\pm$	0.3	\\
N2	&	3.7	&	71.2	$\pm$	8.7	&	96.9	$\pm$	11.8	&	118.0	$\pm$	16.3	&	1.2	$\pm$	0.2	\\
N3	&	3.5	&	73.6	$\pm$	7.9	&	100.1	$\pm$	10.8	&	125.4	$\pm$	5.9	&	1.3	$\pm$	0.1	\\
N4	&	2.2	&	17.7	$\pm$	3.0	&	24.0	$\pm$	4.1	&	58.6	$\pm$	9.6	&	2.4	$\pm$	0.6	\\
N5	&	2.9	&	37.5	$\pm$	5.5	&	51.0	$\pm$	7.4	&	25.1	$\pm$	2.7	&	0.5	$\pm$	0.1	\\
N6	&	2.9	&	33.8	$\pm$	5.6	&	46.0	$\pm$	7.6	&	34.6	$\pm$	5.2	&	0.8	$\pm$	0.2	\\
N7	&	3.4	&	79.1	$\pm$	7.6	&	107.5	$\pm$	10.3	&	79.3	$\pm$	9.9	&	0.7	$\pm$	0.1	\\
N8	&	2.5	&	51.5	$\pm$	4.2	&	70.0	$\pm$	5.7	&	57.5	$\pm$	5.6	&	0.8	$\pm$	0.1	\\
N9	&	3.7	&	94.5	$\pm$	8.8	&	128.5	$\pm$	12.0	&	104.2	$\pm$	14.1	&	0.8	$\pm$	0.1	\\
N10	&	2.8	&	32.7	$\pm$	5.2	&	44.5	$\pm$	7.1	&	34.3	$\pm$	4.1	&	0.8	$\pm$	0.2	\\
N11	&	4.1	&	133.1	$\pm$	11.1	&	181.0	$\pm$	15.0	&	85.8	$\pm$	15.1	&	0.5	$\pm$	0.1	\\
B1	&	3.1	&	48.6	$\pm$	6.3	&	66.1	$\pm$	8.6	&	316.6	$\pm$	71.0	&	4.8	$\pm$	1.2	\\
B2	&	3.3	&	80.0	$\pm$	7.2	&	108.8	$\pm$	9.9	&	267.4	$\pm$	29.8	&	2.5	$\pm$	0.4	\\
B3	&	3.0	&	77.0	$\pm$	5.7	&	104.7	$\pm$	7.7	&	208.8	$\pm$	43.4	&	2.0	$\pm$	0.4	\\
D1	&	3.3	&	56.5	$\pm$	6.9	&	76.9	$\pm$	9.4	&	269.1	$\pm$	62.8	&	3.5	$\pm$	0.9	\\
D2	&	3.5	&	58.6	$\pm$	8.2	&	79.7	$\pm$	11.1	&	207.7	$\pm$	102.0	&	2.6	$\pm$	1.3	\\
D3	&	2.9	&	30.1	$\pm$	5.5	&	41.0	$\pm$	7.4	&	152.3	$\pm$	34.2	&	3.7	$\pm$	1.1	\\
D4	&	5.4	&	117.7	$\pm$	18.9	&	160.0	$\pm$	25.7	&	303.0	$\pm$	47.2	&	1.9	$\pm$	0.4	\\
D5	&	4.1	&	112.6	$\pm$	11.0	&	153.1	$\pm$	15.0	&	474.8	$\pm$	36.8	&	3.1	$\pm$	0.4	\\
\enddata

\tablecomments{
(1) Diameter of the clumps.
(2) H$_{2}$ mass integrated over the diameter of the clumps.
(3) Gas mass integrated over the diameter of the clumps. The
$M_{\rm gas}$ is the $M_{\rm H_2}$ corrected with the He
fraction of 1.36 in Sect.\~ref{sect-mass}.
(4) Virial mass of the clumps.
(5) Ratio of the virial mass to the $M_{\rm gas}$.
We have not corrected for the beam convolution effect as the derived diameters are sufficiently large. We think the assumption of Gaussian shapes may be the greater source of error.
}
\end{deluxetable}

\clearpage

\begin{deluxetable}{cccccc}
\tablewidth{0pt}
\tabletypesize{\scriptsize}
\tablecaption{Star formation properties \label{t.sfr}}

\tablehead{\colhead{ID} & \colhead{$F_{\rm Pa{\alpha}} $} & \colhead{$E$(B-V)}&
\colhead{$A_{\rm Pa\alpha}$}&
\colhead{$A_{\rm v}$} & \colhead{$\Sigma_{\rm SFR}$}\\
\colhead{} & \colhead{(1)} & \colhead{(2)} & \colhead{(3)} & \colhead{(4)} & \colhead{(5)}\\
\colhead{} & \colhead{(10$^{-14}$ erg s$^{-1}$ cm$^{-2}$)} & \colhead{(mag)} & \colhead{(mag)} & \colhead{(mag)} & \colhead{(M$_{\odot}$ yr$^{-1}$ kpc$^{-2}$)}}

\startdata
N1	&	6.03	$\pm$	0.55	&	1.41	$\pm$	0.01	&	0.64	$\pm$	0.30	&	4.36	$\pm$	0.04	&	2.74	$\pm$	0.25	\\
N2	&	7.01	$\pm$	0.39	&	1.06	$\pm$	0.01	&	0.48	$\pm$	0.23	&	3.30	$\pm$	0.04	&	3.18	$\pm$	0.18	\\
N3	&	4.00	$\pm$	0.22	&	1.02	$\pm$	0.02	&	0.46	$\pm$	0.22	&	3.16	$\pm$	0.06	&	1.81	$\pm$	0.10	\\
N4	&	8.95	$\pm$	0.62	&	1.21	$\pm$	0.01	&	0.55	$\pm$	0.26	&	3.75	$\pm$	0.03	&	4.06	$\pm$	0.28	\\
N5	&	7.52	$\pm$	0.38	&	1.00	$\pm$	0.01	&	0.46	$\pm$	0.22	&	3.11	$\pm$	0.03	&	3.41	$\pm$	0.17	\\
N6	&	7.94	$\pm$	0.37	&	0.97	$\pm$	0.01	&	0.44	$\pm$	0.21	&	3.00	$\pm$	0.03	&	3.60	$\pm$	0.17	\\
N7	&	8.97	$\pm$	0.75	&	1.36	$\pm$	0.01	&	0.62	$\pm$	0.29	&	4.20	$\pm$	0.03	&	4.07	$\pm$	0.34	\\
N8	&	4.21	$\pm$	0.41	&	1.45	$\pm$	0.02	&	0.66	$\pm$	0.31	&	4.50	$\pm$	0.06	&	1.91	$\pm$	0.19	\\
N9	&	3.72	$\pm$	0.35	&	1.41	$\pm$	0.02	&	0.64	$\pm$	0.30	&	4.36	$\pm$	0.07	&	1.69	$\pm$	0.16	\\
N10	&	2.90	$\pm$	0.30	&	1.47	$\pm$	0.03	&	0.67	$\pm$	0.32	&	4.56	$\pm$	0.09	&	1.32	$\pm$	0.14	\\
N11	&	1.92	$\pm$	0.22	&	1.49	$\pm$	0.05	&	0.68	$\pm$	0.32	&	4.62	$\pm$	0.14	&	0.87	$\pm$	0.10	\\
B1	&	2.55	$\pm$	0.19	&	1.17	$\pm$	0.03	&	0.53	$\pm$	0.25	&	3.62	$\pm$	0.10	&	1.16	$\pm$	0.09	\\
B2	&	2.60	$\pm$	0.24	&	1.35	$\pm$	0.03	&	0.61	$\pm$	0.29	&	4.18	$\pm$	0.10	&	1.18	$\pm$	0.11	\\
B3	&	1.29	$\pm$	0.17	&	1.44	$\pm$	0.07	&	0.65	$\pm$	0.31	&	4.45	$\pm$	0.21	&	0.58	$\pm$	0.08	\\
D1	&		$\le$0.12		&		-		&		-		&		-		&		-		\\
D2	&		$\le$0.12		&		-		&		-		&		-		&		-		\\
D3	&		$\le$0.12		&		-		&		-		&		-		&		-		\\
D4	&		$\le$0.12		&		-		&		-		&		-		&		-		\\
D5	&	1.56	$\pm$	0.16	&	1.31	$\pm$	0.05	&	0.60	$\pm$	0.28	&	4.07	$\pm$	0.17	&	0.71	$\pm$	0.07		

\enddata

\tablecomments{
(1) Pa$\alpha$ flux corrected by the extinction measured in the
CO clumps. The upper limit of the clumps (D1, ..., D4) is
1.21 $\times10^{-15}$ erg s$^{-1}$  cm$^{-2}$.
(2) Color excess.
(3) Extinction at wavelength of Pa$\alpha$ in the unit of magnitude.
(4) Extinction at V-band in the unit of magnitude.
(5) Surface density of SFR.
The upper limit is 0.05 M$_{\odot}$ yr$^{-1}$ kpc$^{-2}$.}

\end{deluxetable}

\clearpage

\begin{deluxetable}{ccc}
\tablewidth{0pt}
\tabletypesize{\scriptsize}
\tablecaption{Intensity ratios of the CO clumps \label{t.ratio}}

\tablehead{\colhead{ID} & \colhead{R$_{32}$} & \colhead{R$_{13}$} \\
\colhead{} & \colhead{(1)} & \colhead{(2)} }

\startdata
N1	&	1.18 $\pm$ 0.07 (0.78 $\pm$ 0.18)	&	10.52 $\pm$ 7.05		\\
N2	&	0.89 $\pm$ 0.03 (1.02 $\pm$ 0.14)	&	9.17 $\pm$ 2.92	\\
N3	&	1.22	$\pm$ 0.04 &	8.12 $\pm$ 2.35	\\
N4	&	1.41 $\pm$ 0.09	&	-	\\
N5	&	1.68 $\pm$ 0.09	&	12.44 $\pm$ 9.74	\\
N6	&	1.19 $\pm$ 0.07	&	12.47 $\pm$ 9.50\\
N7	&	0.92 $\pm$ 0.02 (0.99 $\pm$ 0.06)	&	8.52 $\pm$ 1.71		\\
N8	&	0.62 $\pm$ 0.02 (0.50 $\pm$ 0.06)	&	10.24 $\pm$ 3.17		\\
N9	&	0.66 $\pm$ 0.03	&	10.88 $\pm$ 5.14		\\
N10	&	0.85 $\pm$ 0.05	&	7.18 $\pm$ 2.73		\\
N11	&	0.78 $\pm$ 0.03	&	13.04 $\pm$ 6.45\\
B1	   &	0.79 $\pm$ 0.03	&	-		\\
B2	   &	0.71 $\pm$ 0.02	&	6.51 $\pm$ 1.10	\\
B3	   &	0.72 $\pm$ 0.02	&	11.49 $\pm$ 3.74		\\
D1	&	0.76 $\pm$ 0.08	&	-		\\
D2	&	0.86 $\pm$ 0.05	&	-	\\
D3	&	1.19 $\pm$ 0.12	&	-	\\
D4	&	0.72 $\pm$ 0.03	&	-		\\
D5	&	0.72 $\pm$ 0.02 (0.39 $\pm$ 0.03)	&	10.65 $\pm$ 2.59 (8.38 $\pm$ 3.41)		\\
Nu	&	0.93 $\pm$ 0.02	&	23.37 $\pm$ 9.79	
\enddata

\tablecomments{
(1) $^{12}$CO(J = 3--2)/(J = 2--1) intensity ratios derived from CO brightness temperature. Numbers within parenthesis are beam smearing corrected ratios.
(2) $^{12}$CO(J = 2--1)/$^{13}$CO(J = 2--1) intensity ratios derived from CO
brightness temperature. Numbers within parenthesis are beam smearing corrected ratios.
All the quantities are measured with a beam size of
3\farcs25$\times$2\farcs55.}

\end{deluxetable}


\begin{deluxetable}{lc}
\tablewidth{0pt}
\tabletypesize{\scriptsize}
\tablecaption{Dynamical parameters fitted by {\tt GAL}\label{t.gal}}
\tablehead{
\colhead{Parameters} &
\colhead{Value}
}

\startdata
R.A. & 02$^{\rm h}$46$^{\rm m}$18$\fs$95\\
Decl. & --30$\degr$16$\arcmin$29\farcs13\\
Position angle & 133\fdg0$\pm0\fdg1$ \\
Inclination & 41\fdg7$\pm0\fdg6$ \\
Systemetic velocity (km s$^{-1}$; $V_{\rm sys}$) & 1249.0$\pm$0.5 \\
$V_{\rm max}$ (km s$^{-1}$) & 387.6$\pm$4.3 \\
$R_{\rm max}$ & 9\farcs5$\pm$0\farcs1
\enddata

\tablecomments{The 6-cm peak of the nucleus is
R.A. = 02$^{\rm h}$46$^{\rm m}$18$\fs$96, Decl. = --30$\degr$16$\arcmin$28\farcs897.
}

\end{deluxetable}


\clearpage

\begin{center}
\begin{figure}
\epsscale{0.25}
\includegraphics[scale=0.48]{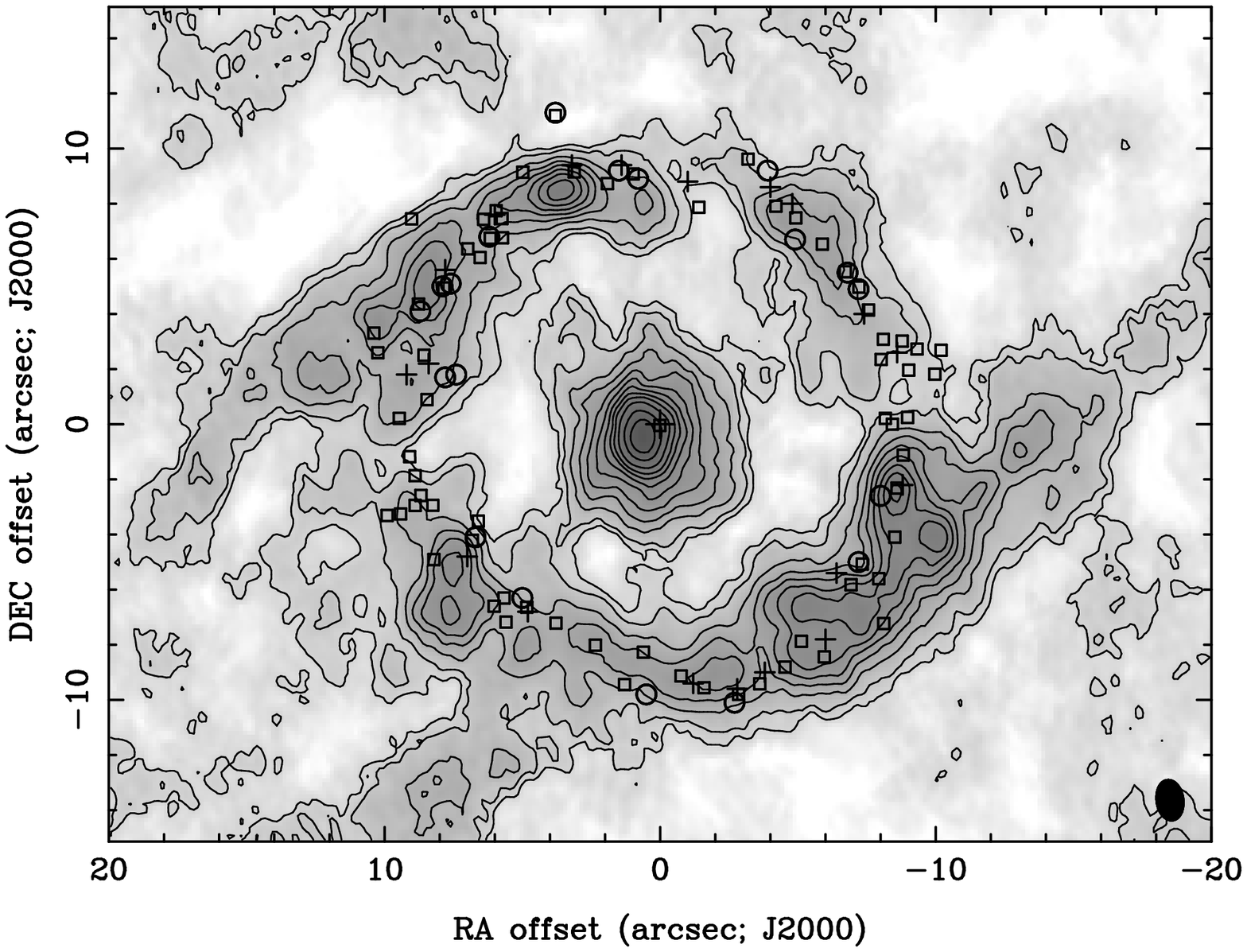}
\includegraphics[scale=0.48]{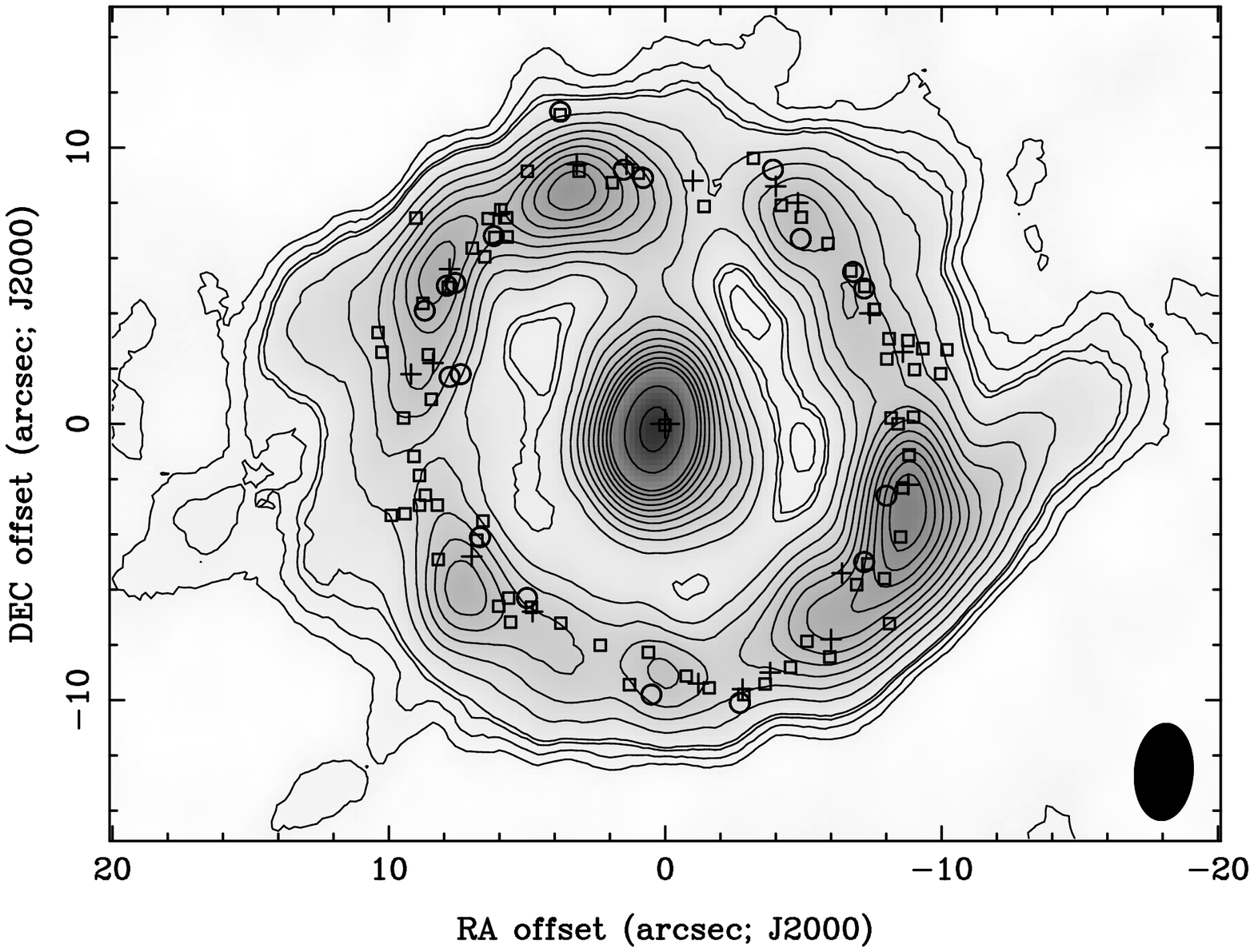}
\includegraphics[scale=0.48]{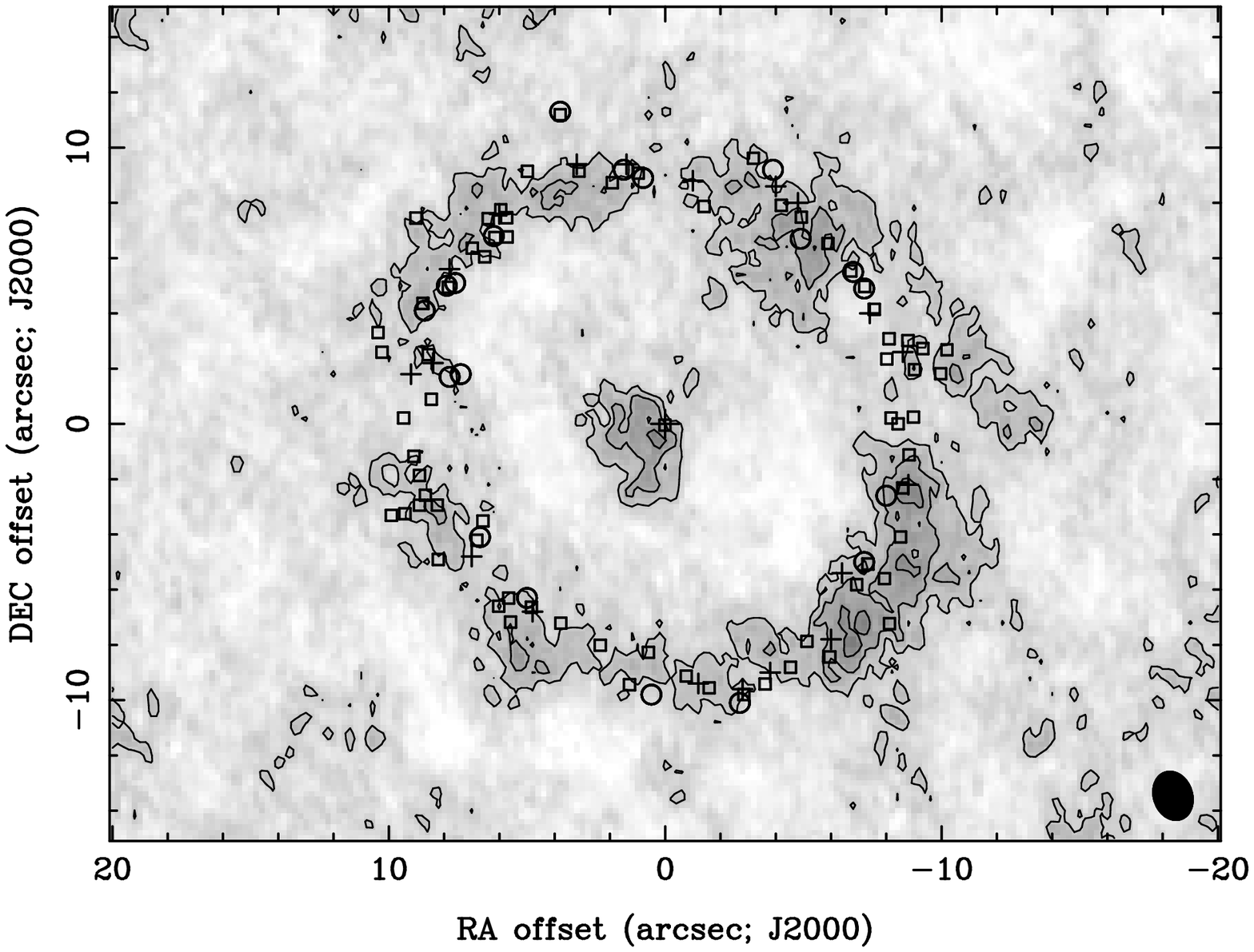}
\includegraphics[scale=0.48]{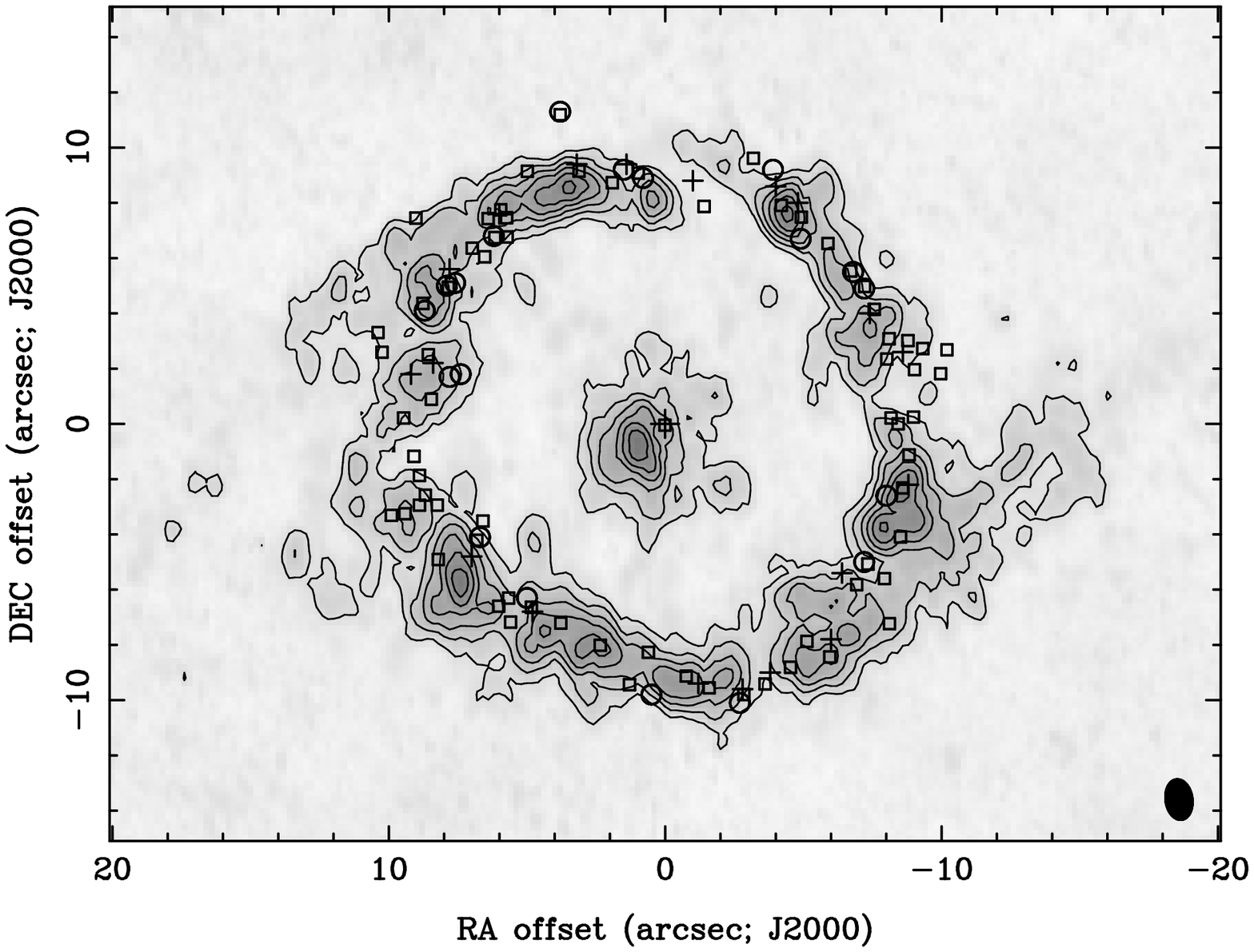}
\scriptsize
\caption[]
{
Top left image is the $^{12}$CO(J = 2--1) integrated intensity map.
The contour levels are 2, 3, 5, ..., 20, 25, and 30$\sigma$
(1$\sigma$  = 2.3 Jy km s$^{-1}$ beam$^{-1}$). The synthesized beam
is 1\farcs5$\times$1\farcs0 (PA = 8\fdg1). The emission
at (10\arcsec,--15\arcsec) are sidelobes.
Top right image is the $^{12}$CO(J = 3--2) integrated intensity map. The
contour levels are 5, 7, 9, 10, 15,..., 60, 80, and 100$\sigma$ (1$\sigma$  = 3.0 Jy km s$^{-1}$ beam$^{-1}$).
The synthesized beam is 3\farcs5$\times$2\farcs1 (PA = --4\fdg4).
Bottom left image is the $^{13}$CO(J = 2--1) integrated intensity map. The contour levels are
2, 3, and 4$\sigma$ (1$\sigma$ = 2.1 Jy km s $^{-1}$ beam$^{-1}$).
The synthesized beam is 1\farcs8$\times$1\farcs4 (PA = 19\degr).
Bottom right image is the $^{12}$CO(J = 2--1) peak brightness temperature map. The contours are 2, 3, 4, 5, 6, and 7 K.
All of the maps are overlaid with the positions of Pa$\alpha$ star clusters (squares),
6 cm radio continuum sources (crosses), and V-band ($<$ 13 mag) star clusters (circles).
The central cross in each map is the position of the 6-cm nucleus (Hummel et al. 1987),
which is assumed to be the active nucleus. The beam size is
shown in the lower right corner of each map.}

\label{fig-mom0}
\end{figure}
\end{center}

\begin{center}
\begin{figure}
\epsscale{1}
\plotone{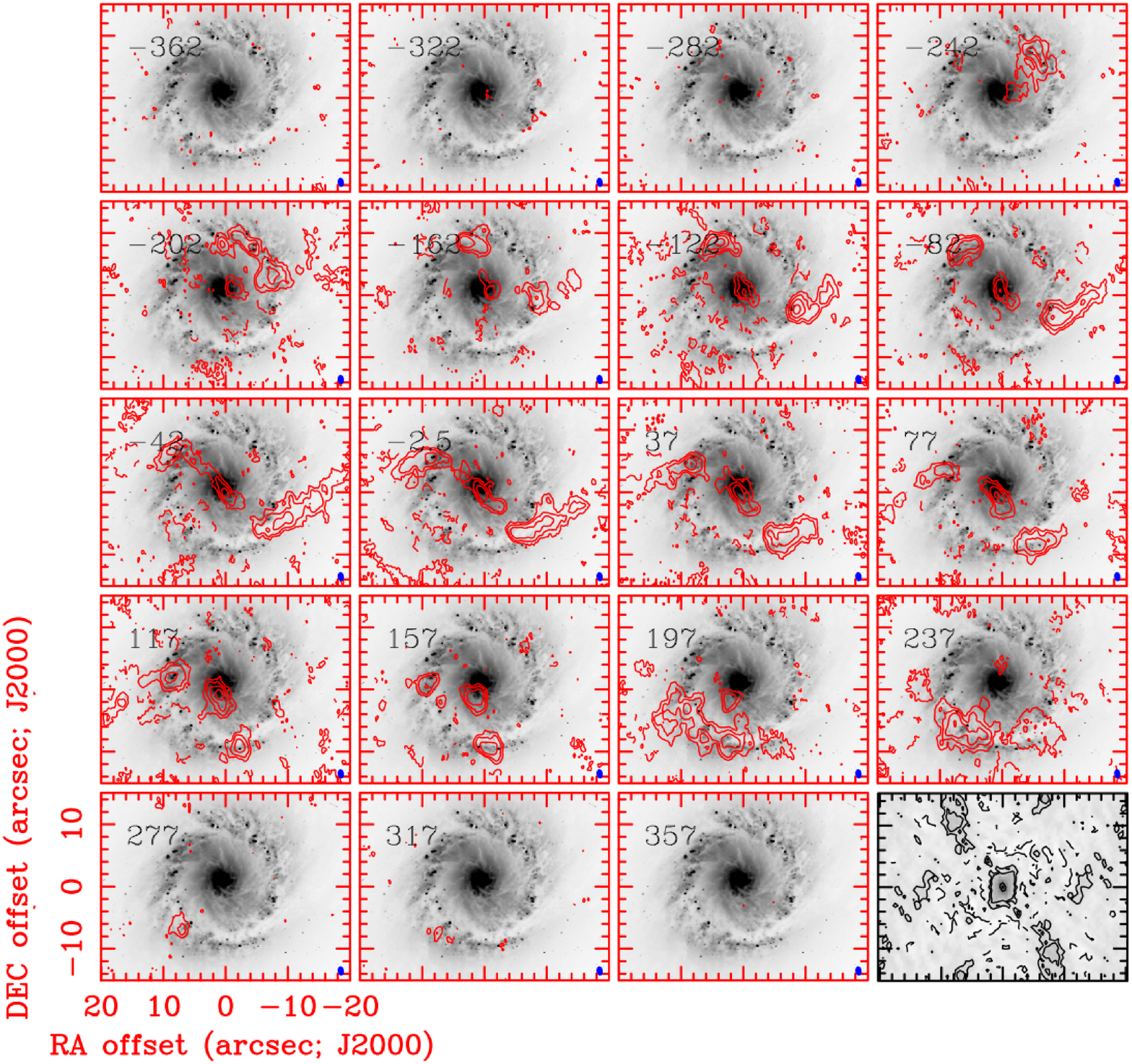}
\caption[]{
The $^{12}$CO(J = 2--1) channel maps are overlaid on the archival HST I-band
(F814W) image with the corrected astrometry.
The contour levels are  --2, 2, 4, 8, 16, and 32$\sigma$, where 1$\sigma$ = 20 mJy
beam$^{-1}$ (306 mK) in 40 km s$^{-1}$ resolution.
The velocity (km s$^{-1}$) with respect to the systemic velocity
of 1254 km s$^{-1}$ (Kohno et al. 2003) is labeled in the top left corner of each map.
The beam size (1\farcs5$\times$1\farcs0, PA = 8\fdg1) is shown in the lower
right corner of each map with solid ellipse. The dirty beam is at the bottom right panel
with a contours level of --100, --50, --10, --5, 5, 10, 50, 100\% of the peak.}
\label{fig-channel1}
\end{figure}
\end{center}

\begin{figure}
\epsscale{1}
\plotone{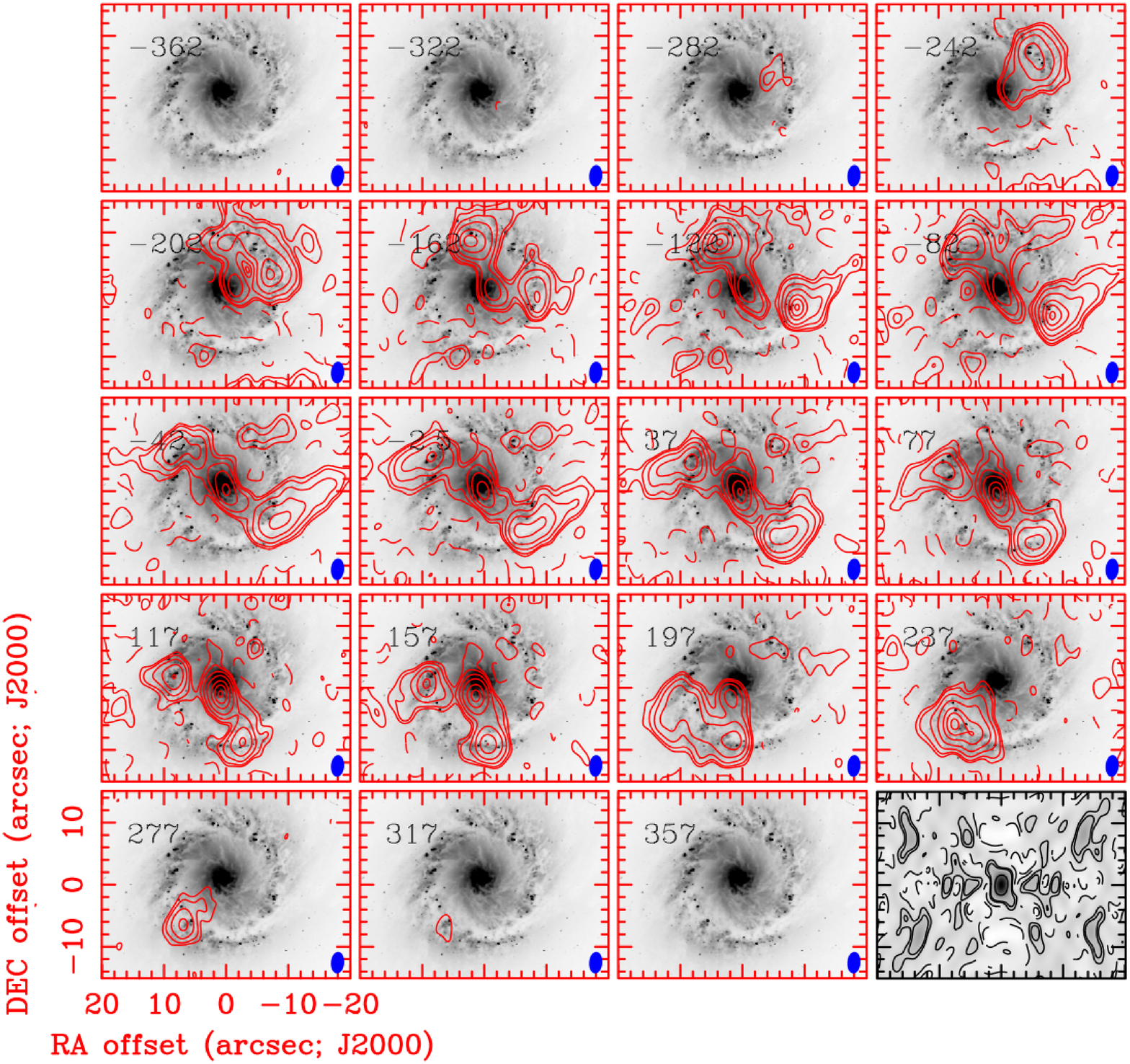}
\caption[]{
The $^{12}$CO(J = 3--2) channel maps are overlaid on the archival HST I-band
(F814W) image with the corrected astrometry.
The contour levels are -3, 3, 5, 10, 20, 40, 60, and 80$\sigma$, where
1$\sigma$ = 17.5 mJy beam$^{-1}$ (23 mK) in 40 km s$^{-1}$ resolution.
The velocity (km s$^{-1}$) with respect to the systemic velocity
of 1254 km s$^{-1}$ is labeled in the top left corner of each map.
The beam size (3\farcs5$\times$2\farcs1, PA = --4\fdg4) is shown in the lower
right corner with solid ellipse. The dirty beam is at the bottom right panel
with a contours level of --100, --50, --10, --5, 5, 10, 50, 100\% of the peak.}
\label{fig-channel2}
\end{figure}

\begin{center}
\begin{figure}
\epsscale{0.5}
\includegraphics[scale=0.35,angle=-90]{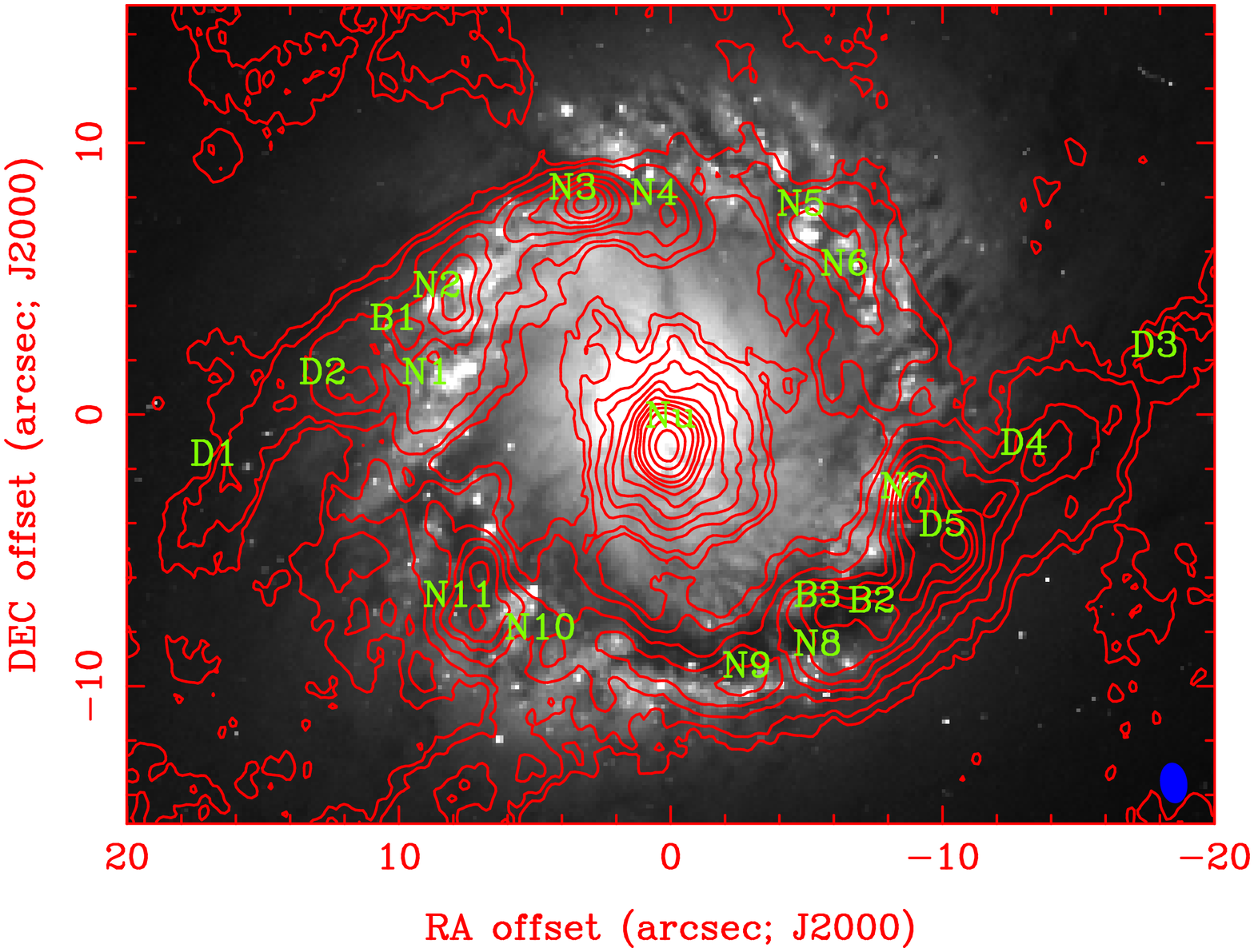}
\includegraphics[scale=0.35,angle=-90]{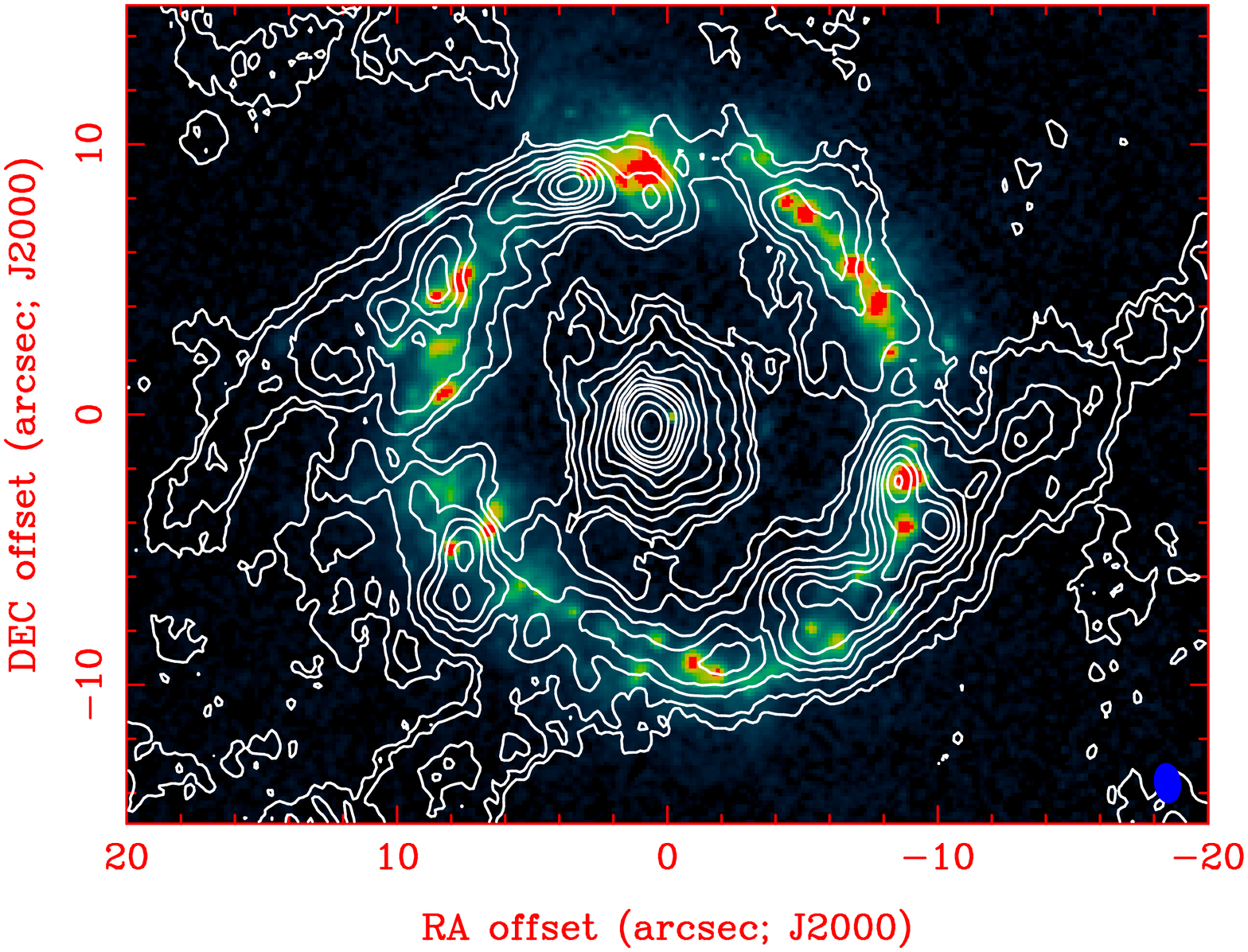}
\caption[]{
Top image is the $^{12}$CO(J = 2--1) integrated map (contours) overlaid on the archival
HST I-band (Filter F814 W) image (grey scale). Astrometry of the HST
image was corrected using background stars with known positions.
The contour levels for the $^{12}$CO(J = 2--1) are 2, 3, 5, ..., 20, 25, and 30 $\sigma$
(1 $\sigma$ = 2.3 Jy km s$^{-1}$ beam$^{-1}$). The IDs for the individual
peaks of clumps are marked. The CO synthesized beam (1\farcs5$\times$
1\farcs0) is shown in the lower right corner.
Bottom image is the HST NICMOS Pa$\alpha$ line image (color) overlaid on the $^{12}$CO(J = 2--1) contour. The contour levels are the same as in upper image.}
\label{fig-mom0-hst}
\end{figure}
\end{center}

\begin{center}
\begin{figure}
\epsscale{1}
\plotone{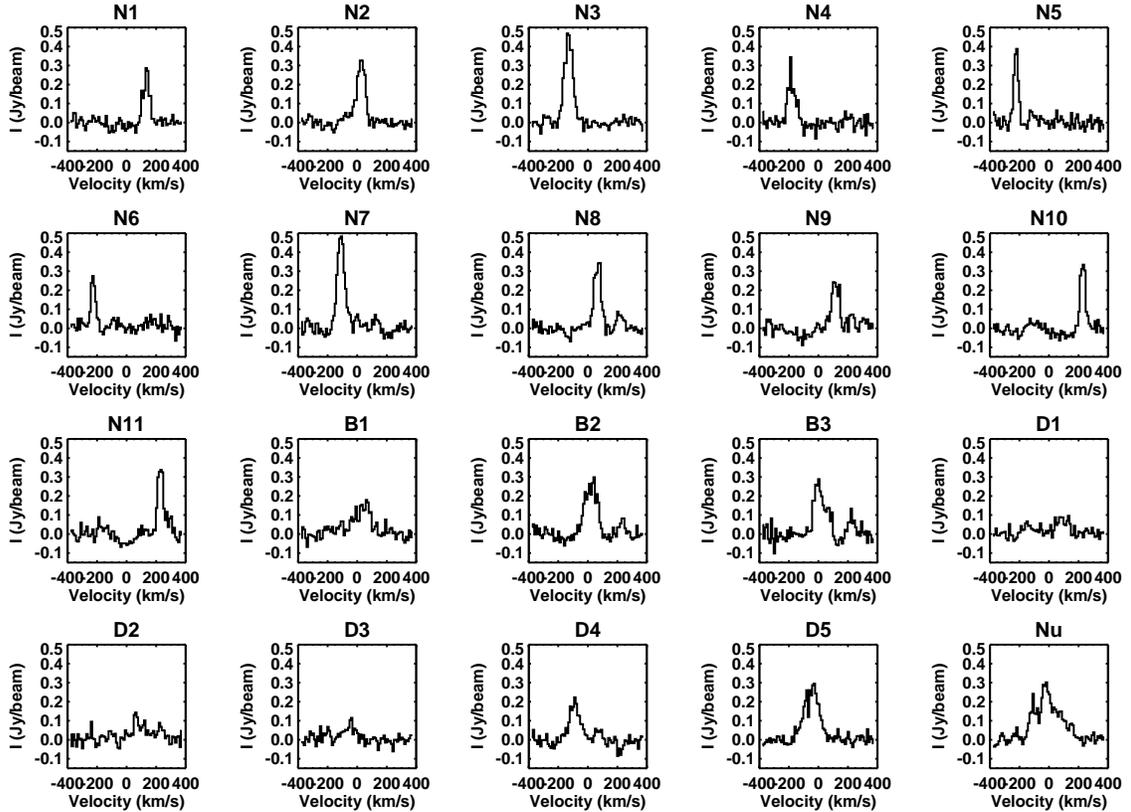}
\caption[]{
The high resolution (1\farcs5$\times$1\farcs0) $^{12}$CO(J = 2--1)
spectra of individual clumps measured at the $^{12}$CO(J = 2--1) peak
position within one beam. The velocity is relative to the systemic velocity of 1254 km s$^{-1}$ and spectral resolution is 10 km s$^{-1}$. The IDs of
the clumps are labeled in each panel.}
\label{fig-spectra}
\end{figure}
\end{center}

\begin{center}
\begin{figure}
\epsscale{1}
\plotone{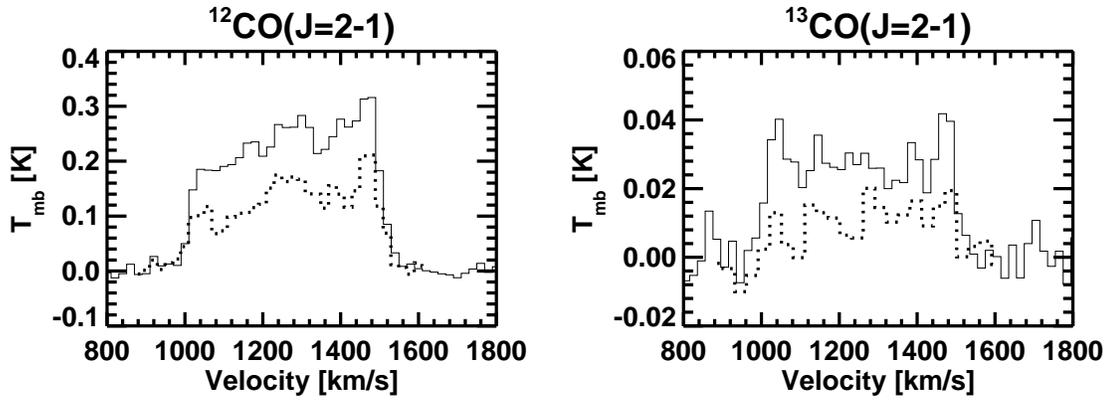}
\caption[]{
Left: We show the $^{12}$CO(J = 2--1) spectra, where
the solid line is the JCMT data (Petitpas et al. 2003) and the dotted
line is our SMA data. The intensity scale is the main beam temperature
at 21\arcsec~resolution. We only took 6 chunks to make
the SMA maps, so the velocity range is smaller
than that of JCMT.
Right: The $^{13}$CO(J = 2--1) spectra, where
the solid line is the JCMT data, and the dotted line is our SMA data.
The beam size of the two data are matched to 21$\arcsec$.}

\label{fig-single}
\end{figure}
\end{center}

\begin{center}
\begin{figure}
\epsscale{1}
\plotone{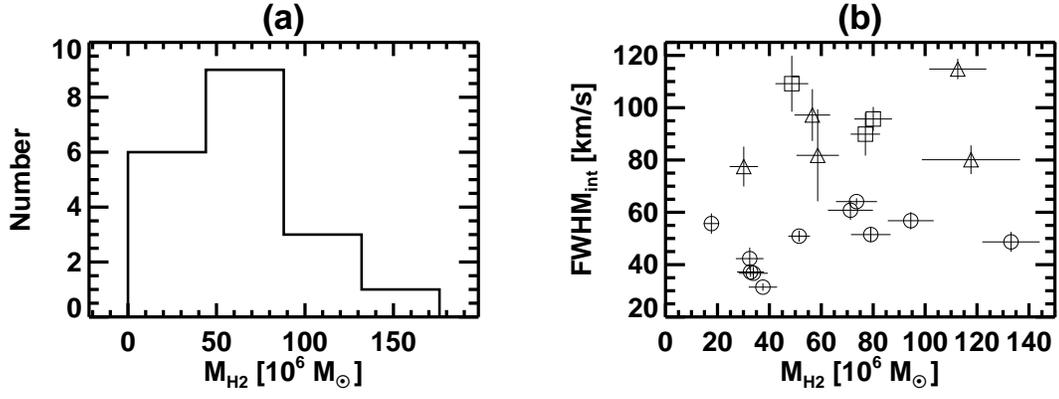}
\caption[]{
(a) The number histogram of the total H$_{2}$ mass of the clumps in the ring.
Horizontal and vertical axis are the gas mass and number, respectively.
The gas mass is in units of 10$^{6}$ M$_{\odot}$. The negative horizontal
axis is to show the plot clearly.
(b) The correlation between total H$_{2}$ mass and FWHM intrinsic
line width of the narrow line ring clumps (circles), broad line ring
clumps (squares), and dust lane clumps (triangles).
The H$_{2}$ mass is in units of 10$^{6}$ M$_{\odot}$.
The uncertainties of $\pm1\sigma$ are overlaid on the symbols
with vertical/horizontal bars.}
\label{fig-clump-prop}
\end{figure}
\end{center}

\begin{center}
\begin{figure}
\epsscale{1}
\plotone{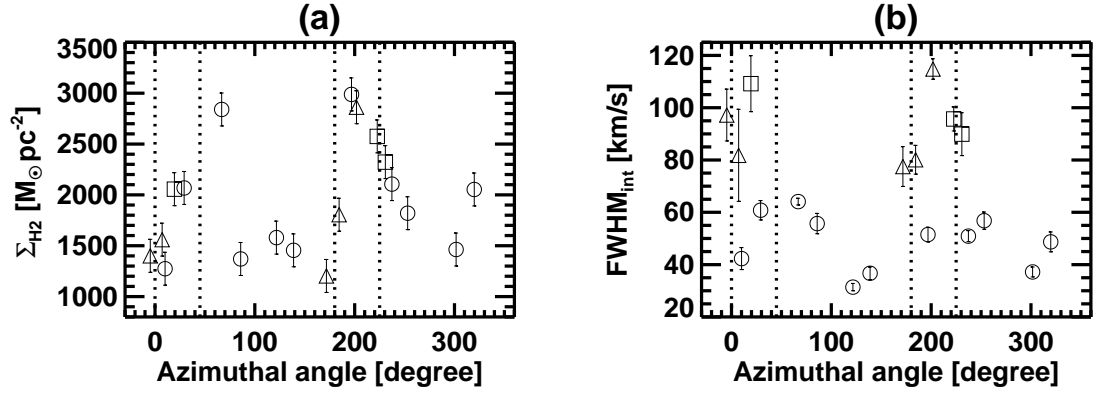}
\caption[]{
(a) $\Sigma_{\rm H_{2}}$ measured at the position of the intensity
peak in units of M$_{\odot}$ pc$^{-2}$ (see Table~\ref{t.clumps}) as a function of azimuthal angle. East direction corresponds to 0\degr, and increase is in clockwise direction.
The dashed lines mark position angle from 0\degr~to 45\degr and
from 180\degr~to 225\degr,
which roughly correspond to the position of the orbit crowding regions.
The meaning of the symbols are the same as in Figure~\ref{fig-clump-prop}b.
(b) The FWHM intrinsic line width of clumps as a function of
azimuthal angle. The dashed lines are the same as in (a).
The uncertainties of $\pm1\sigma$ are overlaid on the symbols
with vertical/horizontal bars.}
\label{fig-clump-peak}
\end{figure}
\end{center}

\begin{center}
\begin{figure}
\epsscale{1}
\plotone{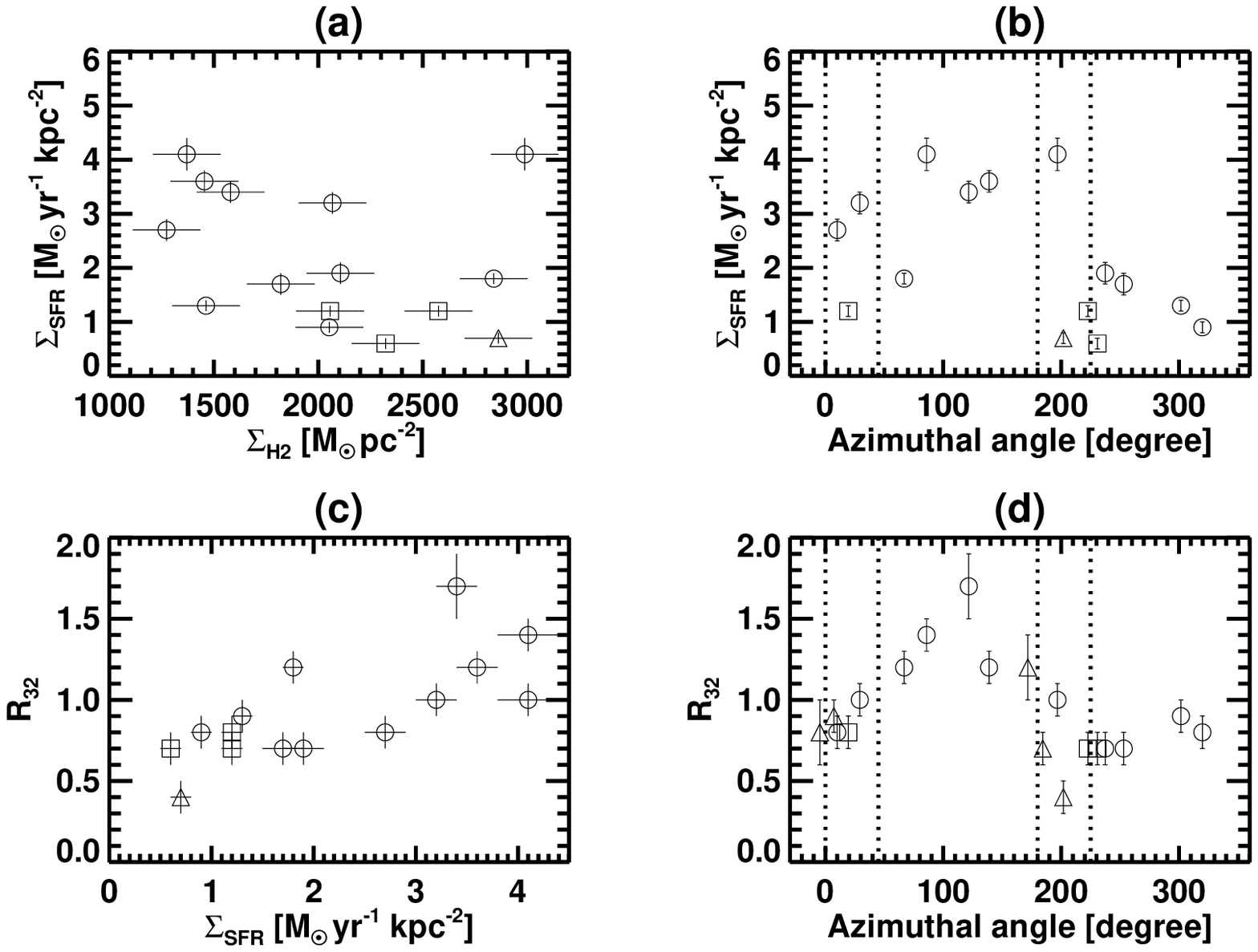}
\caption[]{
(a) Surface SFR density (M$_{\odot}$ yr$^{-1}$ kpc$^{-2}$) is shown as
a function of $\Sigma_{\rm H_{2}}$ (M$_{\odot}$ pc$^{-2}$).
The symbols are the same as in Figure~\ref{fig-clump-peak}.
The meaning of the symbols are the same as in Figure~\ref{fig-clump-prop}b.
(b) Surface SFR density is shown as a function of azimuthal angle.
The dashed lines mark position angle from 0\degr~to 45\degr and
from 180\degr~to 225\degr.
(c) Surface SFR density (M$_{\odot}$ yr$^{-1}$ kpc$^{-2}$)
is shown as a correlation of R$_{32}$.
(d) R$_{32}$ is shown as a function of azimuthal angle.
The uncertainties of $\pm1\sigma$ are overlaid on the symbols
with vertical/horizontal bars.}
\label{fig-clump-sfr}
\end{figure}
\end{center}

\begin{figure}
\epsscale{0.85}
\plotone{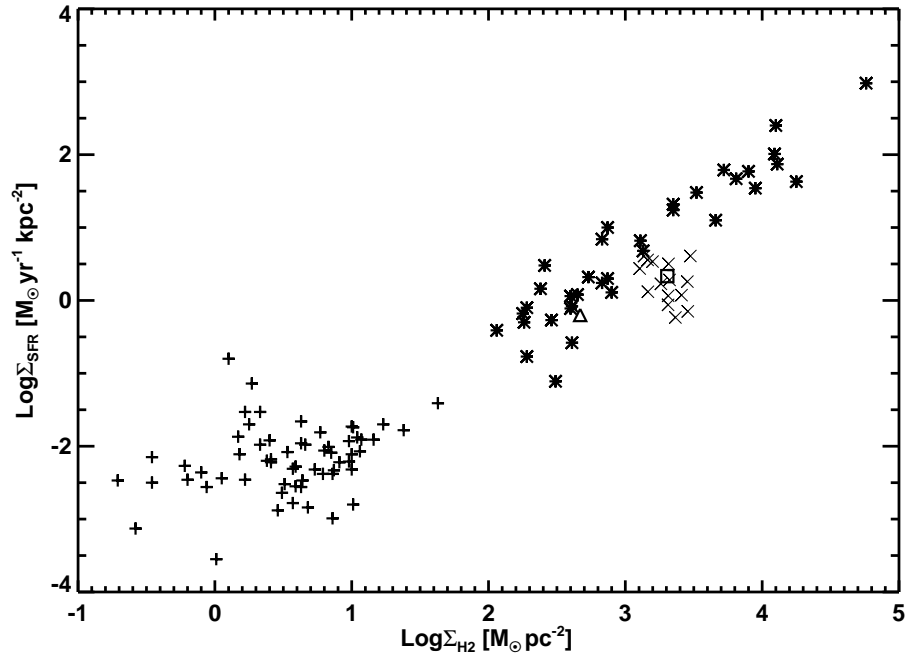}
\caption[]{The correlation of $\Sigma_{\rm H_{2}}$ and surface
SFR density of individual clumps of NGC 1097
are overlaid on the data used in Kennicutt (1998).
Their data for normal galaxies are represented as plus signs, and
infrared-selected
circumnuclear starburst galaxies as asterisks, while their NGC 1097
data point is marked as triangle.
Our spatially resolved clumps of the circumnuclear starburst ring
of NGC 1097 are marked as crosses, and
the average value of the clumps is marked as a square.}
\label{fig-kslaw}
\end{figure}

\begin{center}
\begin{figure}
\epsscale{1}
\plotone{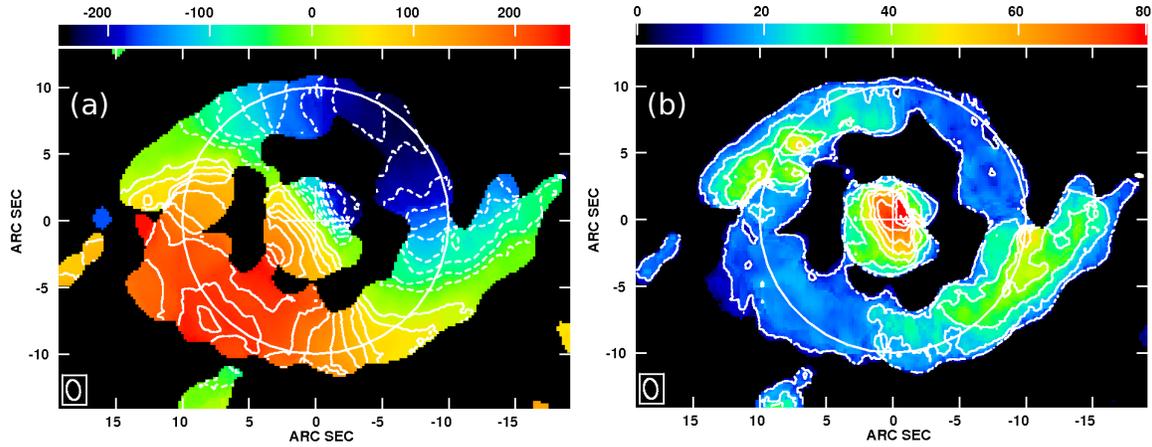}
\caption[]{(a) The intensity weighted mean velocity
map (MOM1) of $^{12}$CO(J = 2--1) line
with respect to the systematic velocity (1254 km s$^{-1}$), solid and
dashed lines represent the redshifted and blueshifted velocity
respectively. The first negative contour (close to the central
cross) is 0 km s$^{-1}$, and the contour spacing is in
25 km s$^{-1}$ resolution.
(b) The intensity weighted velocity dispersion map (MOM2).
The contour interval is 10 km s$^{-1}$, note the values are
not FWHM line width but the square root of the dispersion relative to the mean velocity.
Therefore the number is lower than the FWHM we derived by line fitting.}
\label{fig-mom1}
\end{figure}
\end{center}

\begin{center}
\begin{figure}
\epsscale{0.8}
\plotone{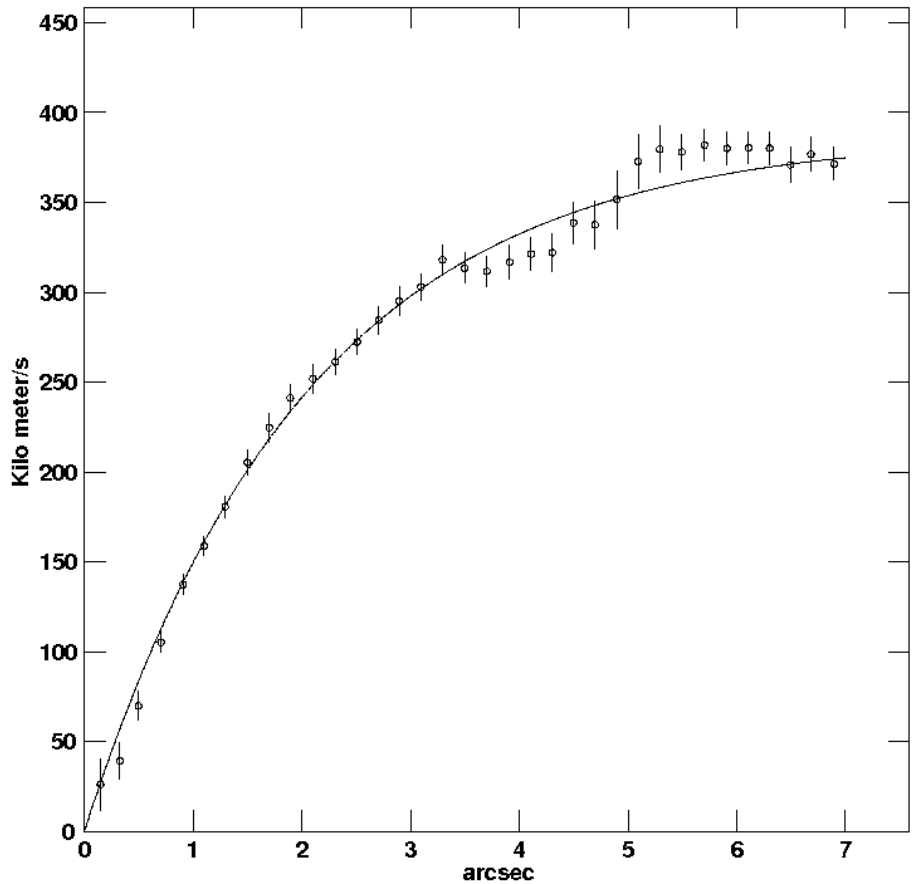}
\caption[]{
We show the data of rotation curve (circles) of NGC 1097 overlaid with the fitted
curve (solid line) in {\tt GAL}.}
\label{fig-gal}
\end{figure}
\end{center}

\begin{center}
\begin{figure}
\epsscale{0.7}
\plotone{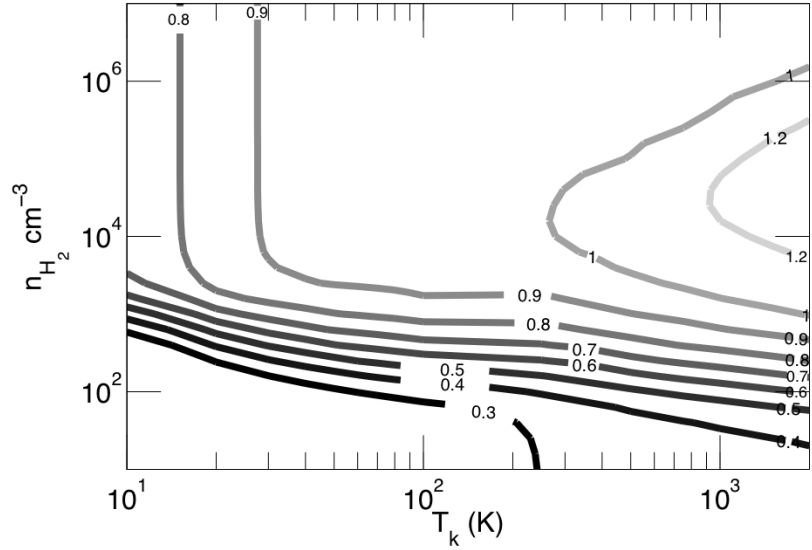}
\caption[]{The LVG calculations of the R$_{32}$ is shown
as a function of kinetic temperature and H$_{2}$ number density.
The R$_{32}$ are labeled on the contours.
}
\label{fig-lvg}
\end{figure}
\end{center}

\begin{center}
\begin{figure}
\epsscale{1}
\plotone{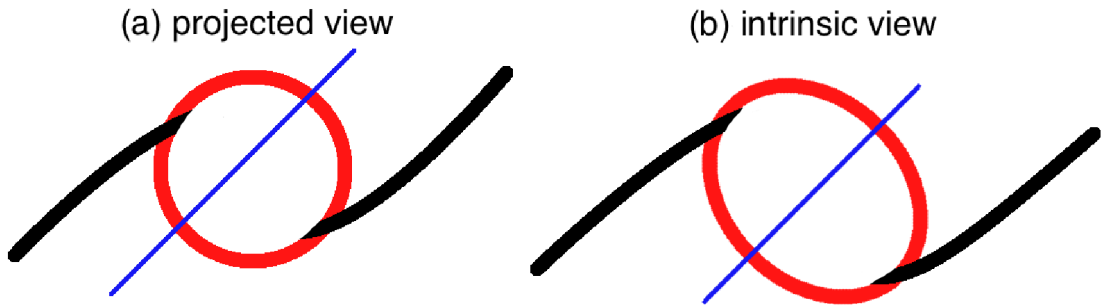}
\caption[]{The cartoon sketch of the gas morphology in the circumnuclear
region of NGC 1097. The
red circle represents the starburst ring, where the narrow line clumps
are located. The black curves are the dust lane associated with shock
wave, where the broad line clumps are located. The blue line is the
major axis of the large scale stellar bar. We show:
(a) the projected view of the morphology of the starburst ring/dust lane
associated with our observation. It shows a nearly circular starburst ring.
(b) the intrinsic shape of the ring, which is expected to be an ellipse.}
\label{fig-model}
\end{figure}
\end{center}

\clearpage

\end{document}